\documentclass[a4paper,11pt]{article}

\usepackage{geometry}										
\usepackage[english]{babel}									
\usepackage[utf8]{inputenc}									
\usepackage[T1]{fontenc}									

\usepackage{lmodern}										

\usepackage{amssymb,amsfonts,amsmath,amsthm}
\usepackage{hyphenat}
\usepackage{booktabs, multirow}
\usepackage{bbm}
\usepackage{graphicx}
\usepackage{algpseudocode}
\usepackage{algorithm}
\usepackage{subfigure}
\usepackage{float} 

\graphicspath{{Figures/}}

\makeatletter
\let\oldabs\abs
\def\abs{\@ifstar{\oldabs}{\oldabs*}}
\let\oldnorm\norm
\def\norm{\@ifstar{\oldnorm}{\oldnorm*}}
\makeatother

\newcommand{\ve}[1]{\boldsymbol{#1}}
\newcommand{\ua}{\ve{\alpha}}
\newcommand{\ca}{{\mathcal A}}
\newcommand{\cm}{{\mathcal M}}
\newcommand{\cn}{{\mathcal N}}
\newcommand{\cx}{{\mathcal X}}
\newcommand{\cy}{{\mathcal Y}}
\newcommand{\Pro}{\mathbb{P}}
\newcommand{\Esp}[1]{{\mathbb E}\left[ #1 \right]} 

\usepackage{authblk}										

\title{Uncertainty-aware multi-fidelity surrogate modeling with noisy data}
\author[1]{Katerina Giannoukou \thanks{katerina.giannoukou@ibk.baug.ethz.ch}}
\author[1]{Stefano Marelli\thanks{marelli@ibk.baug.ethz.ch}}
\author[1]{Bruno Sudret\thanks{sudret@ethz.ch}}
\affil[1]{Chair of Risk, Safety and Uncertainty Quantification, ETH Z\"{u}rich, Switzerland}

\date{\today}

\usepackage{microtype}										

\usepackage{indentfirst}									

\usepackage{caption}
\usepackage{subcaption}

\usepackage{natbib}
\usepackage[hidelinks, colorlinks=true, linkcolor=black, citecolor=black]{hyperref}
\hypersetup{
pdftitle = {Uncertainty-aware multi-fidelity surrogate modeling with noisy data},
pdfauthor = {Katerina Giannoukou, Stefano Marelli, Bruno Sudret},
pdfsubject = {},
pdfkeywords = {multi-fidelity, surrogate modeling, confidence intervals, prediction intervals, polynomial chaos expansions, bootstrap},
colorlinks = false,                                                                                                     
}

\usepackage[capitalize]{cleveref}								
\creflabelformat{equation}{#2(#1)#3}								
\crefrangelabelformat{equation}{#3(#1)#4 to #5(#2)#6}						
\labelcrefformat{subequation}{#2(#1)#3}								
\labelcrefrangeformat{subequation}{#3(#1)#4 to #5(#2)#6}					

\graphicspath{{./figures}} 

\begin{document}

\maketitle

\begin{abstract}
Emulating high-accuracy computationally expensive models is crucial for tasks requiring numerous model evaluations, such as uncertainty quantification and optimization. When lower-fidelity models are available, they can be used to improve the predictions of high-fidelity models.
Multi-fidelity surrogate models combine information from sources of varying fidelities to construct an efficient surrogate model. 
However, in real-world applications, uncertainty is present in both high- and low-fidelity models due to measurement or numerical noise, as well as lack of knowledge due to the limited experimental design budget.
This paper introduces a comprehensive framework for multi-fidelity surrogate modeling that handles noise-contaminated data and is able to estimate the underlying noise-free high-fidelity model. Our methodology quantitatively incorporates the different types of uncertainty affecting the problem and emphasizes on delivering precise estimates of the uncertainty in its predictions both with respect to the underlying high-fidelity model and unseen noise-contaminated high-fidelity observations, presented through confidence and prediction intervals, respectively.
Additionally, the proposed framework offers a natural approach to combining physical experiments and computational models by treating noisy experimental data as high-fidelity sources and white-box computational models as their low-fidelity counterparts.
The effectiveness of our methodology is showcased through synthetic examples and a wind turbine application.
\end{abstract}

\section{Introduction}
Predicting the behavior of complex systems and quantifying the corresponding uncertainty is a ubiquitous challenge in engineering and applied sciences. 
To address this challenge, a variety of predictive models are employed. 
White-box computational models, typically implemented as computer simulations, use pre-existing knowledge on the underlying physics of a system to predict its behaviour. 
Conversely, black-box data-driven models  act as global approximators of the response of a system, based on an available set of input-output observations \citep{Rogers2017}. 
In applications such as uncertainty quantification and optimization, which require numerous model evaluations, a specific subset of black-box models, known as \textit{surrogate models} (SMs)—also known as \textit{metamodels} or emulators—is often used to replace computationally expensive computational models.
A SM acts as an inexpensive-to-evaluate approximator of an original model, and is constructed using a limited set of model evaluations, called the \textit{experimental design} (ED), also known as \textit{training set} in machine learning. 
Among the most widely used SMs for deterministic simulators are polynomial chaos expansions (PCE) \citep{Xiu2002,BlatmanJCP2011}, Gaussian processes (GPs) \citep{Rasmussen2006}, and support vector regression \citep{Drucker1996}. 

A specific class of surrogate models that is particularly useful in scenarios where data or computational models of varying fidelities are accessible is \textit{multi-fidelity surrogate models} (MFSMs). 
With model fidelity, we refer to the extent to which a model faithfully reflects the characteristics and behavior of the target system it intends to simulate. 
Generally, high-fidelity (HF) models produce accurate predictions, but are associated with high computational or financial costs. Low-fidelity models (LF) are instead less accurate, but also less expensive to run.
Models of different fidelities can occur by, e.g., changing the mathematical or numerical model, or changing accuracy of the numerical solver using different levels of discretization \citep{GiselleFernandezGodino2023}.
Multi-fidelity (MF) surrogate modeling approaches combine multiple sources of different fidelity into a single surrogate model, usually augmenting a limited  and expensive-to-obtain HF data set with more extensive and less expensive lower-fidelity ones \citep{kennedy2000,le2014recursive}. 

The choice of surrogate model is an integral part of the design and construction of a MFSM.
Numerous MFSM techniques are based on Gaussian process modeling, following the autoregressive fusing scheme proposed by \citet{kennedy2000}. 
Such works include \citet{Forrester2007} and \citet{le2014recursive}, among others, with the latter reformulating the approach from \citet{kennedy2000} to have a recursive form, allowing for a reduced computational complexity. 
Polynomial chaos expansion is another SM that has gained popularity in the past two decades for the purpose of MF surrogate modeling \citep{Ng2012,Palar2016}. 
The approaches mentioned so far use linear information fusion, which entail assuming that a higher-fidelity response can be expressed as a linear combination of a lower-fidelity model and a discrepancy function.
Recently, multi-fidelity modeling approaches have been proposed in the machine learning community, for example, the deep GP-based framework of \citet{Cutajar2018,Hebbal2021}, the Bayesian neural network approaches from \citet{Meng2021,Kerleguer_2024}, and the generative adversarial network-based methodology proposed by \citet{Zhang2022}. 
These approaches can capture the nonlinear relations between the different levels of fidelity. 
According to a comparison among different linear and nonlinear GP-based MF techniques performed by \citet{Brevault2020a}, when the high- and low-fidelity models are weakly correlated, nonlinear techniques can outperform linear and less complex techniques, with the caveat of requiring a larger quantity of HF data. 

However, multi-fidelity methods do not necessarily involve the construction of a MFSM. 
\citet{Peherstorfer2018} classify MF methods into three categories: \textit{adaptation}, where the LF model is enhanced with information from the HF model; \textit{fusion}, which combines information from all fidelity levels simultaneously; and \textit{filtering}, where the HF model is invoked only when the LF model is inaccurate or some criterion is met. 
Typically, multi-fidelity surrogate modeling methods fall under the adaptation \citep{kennedy2000,Ng2012,Chakraborty2021} or fusion \citep{Bryson2017,Yang2019} categories. In the context of uncertainty quantification, non-surrogate-based MF approaches include the control variate framework \citep{Lavenberg1981,Gorodetsky2020,Pham2022}, which falls under the fusion category, as well as importance sampling \citep{Peherstorfer2016} and multistage Markov Chain Monte Carlo methods \citep{Christen2005}, both classified under filtering.
Moreover, in some MF settings, a strict hierarchy of fidelity levels is not present. The MFNets framework proposed by \citet{Gorodetsky2020MFnets,Gorodetsky2021} provides a general approach to incorporate multiple fidelities with diverse relations.

Most of the existing literature on MF surrogate modeling focuses on deterministic and noise-free high- and low-fidelity models.
However, all real-world measurement devices have limited precision and resolution, and therefore, data resulting from measurements are generally contaminated by irreducible noise. In our context, this noise is considered as a source of aleatory uncertainty, as we assume that no more precise measurement device is available once a specific experimental data set has been provided.
Moreover, the available training data for the construction of all the elements of a MFSM is in principle relatively small due to computational budget constraints. 
Thus, we consider all MFSM predictions as affected by epistemic uncertainty.
Recent studies that consider the presence of noise in a MF setting include the work of \citet{Raissi2017}, who use GP regression to infer the solutions of differential equations when noisy data of different fidelities are available. 
Furthermore, \citet{Zhang2018} demonstrate that their linear regression-based MF surrogate modeling technique is robust to noise in the HF data, and is also able to estimate the noise level when enough HF data is available.
Also, \citet{Ficini2021} assess the robustness of a GP regression MFSM on problems affected by noisy objective function evaluations. 
Finally, few regression-based approaches have combined physical experiments and computer simulations using  MFSMs \citep{Kuya2011,Pandita2021}.

To the authors' best knowledge, however, no work has yet introduced a comprehensive approach to multi-fidelity surrogate modeling that considers the presence of both noise and/or epistemic uncertainty in the high- and/or low-fidelity data. 
This approach should also provide a way to quantify the accuracy of the MF model predictions with respect to a) the underlying noise-free HF model and b) the noisy HF observations. 
The uncertainty about the mean prediction of a regression model is typically expressed via \textit{confidence intervals} (CIs). 
The CIs express the epistemic uncertainty in the MFSM predictions.
On the other hand, the uncertainty on the prediction of an unseen noise-contaminated observation is shown via \textit{prediction intervals} (PIs) \citep{Kutner2005}. PIs consider the total of epistemic and aleatory uncertainty in the MFSM predictions, and thus, generally, they are wider than the corresponding CIs. 

Many of the existing MF surrogate modeling methods do not provide uncertainty estimations about the MFSM predictions. 
When CIs are provided, their construction is usually linked to the particular methodology employed in constructing the MFSM, e.g., GP-based CIs \citep{Raissi2017,Perdikaris2017}. This can potentially make the frameworks less flexible when different MFSM architectures or SMs need to be explored. 
Additionally, GP-based confidence intervals often assume that the response distribution at each point is Gaussian, leading to symmetrical CIs around the MFSM response. However, this assumption does not always reflect reality. We propose a bootstrap-based approach for constructing CIs that does not assume any specific shape for the response distribution, making it more general and applicable to a wider range of scenarios. Moreover, no work has considered the distinction between confidence and prediction intervals in a multi-fidelity setting. Typically, prediction intervals are either disregarded or, mistakenly, confidence intervals are presented as prediction intervals.

The goal of the present paper is to introduce a novel general framework for multi-fidelity surrogate modeling that is able to deal with noisy data and epistemic uncertainty due to limited training information. 
We assume that the noisy HF data originate from deterministic models contaminated by unbiased stochastic noise. Therefore, our MF framework aims at effectively emulating the underlying deterministic models, or in other words at denoising the noise-contaminated observations.
An essential and original feature of our methodology is its ability to provide estimates of the different kinds of uncertainty in its predictions in the form of both confidence and prediction intervals.

Often, the response of a system can be predicted by one or more white-box computational models, while additional data can be obtained through physical experiments. The proposed multi-fidelity framework provides a direct approach for integrating both computational models and experimental data into a single surrogate model.
Traditionally, physical experiments have been used to improve computational models through model calibration, where parameters of a model are inferred by fitting the model predictions to available experimental data \citep{Kennedy2001,Higdon2004,Pavlak2014}. 
However, computational models may not fully capture the system's complexity, making it important to incorporate data-driven elements.
In this paper, we achieve this integration using multi-fidelity surrogate models.
The computational models are assumed to capture the general physical behavior of the system, while the experiments can capture its entirety, but only on a very limited set of conditions, due to their associated costs. 
In particular, we consider the experimental data as noise-contaminated realizations of unknown HF models, whereas the available white-box computational models are considered as their LF counterparts. 
We approach white-box models within a non-intrusive context, eliminating the requirement for prior knowledge of the underlying equations. 
Other methods that have been explored in the literature to combine physical experiments and simulations include hybrid simulation, which combines physical and numerical substructures to create a hybrid model \citep{Schellenberg2009,Abbiati2021}, as well as scientific machine learning approaches, such as physics-informed neural networks \citep{Raissi2019,Jagtap2022} and physics-informed PCE \citep{Novak2024}, which incorporate physical laws and constraints directly into the surrogate models.

This paper is organized as follows: First, we recall the relevant theory for MFSMs and present our framework for MF surrogate modeling, including the special case of combining physical experiments and simulations. We then present the related confidence and prediction intervals, as well as the implementation details for each component of the framework. 
Subsequently, we assess the proposed method by applying it to two synthetic examples of increasing complexity, as well as a real-world application. Finally, we discuss concluding remarks and present prospects for future research.

\section{Methods}
\label{sec:methods}
In this section, we first formally state the multi-fidelity surrogate modeling problem, while establishing the notation we are adopting throughout the paper. 
Then, we describe our proposed methodology to combine experimental data and computational models in a multi-fidelity surrogate model.
Subsequently, we propose to express the uncertainty related to the MFSM predictions using confidence and prediction intervals, and we discuss the interpretation of these intervals in the MF setting.
Lastly, we outline our proposed implementation for constructing a MFSM, as well as for estimating its confidence and prediction intervals.

\subsection{Multi-fidelity surrogate modeling}
We assume that we have $s$ information sources which produce data of different fidelity levels. Without loss of generality, we initially focus on the case when two levels of fidelity are present. 
Let us consider a HF data set $(\ve\cx_\text{H}, \cy_\text{H})$ of size $N_\text H$, obtained, e.g., from an expensive experimental campaign or computational model. 
The input space is $\ve{X} \in \mathbb{R}^M$, while the output space is $Y \in \mathbb{R}$.
We assume that observations from this HF information source are contaminated by additive noise, which can correspond to measurement noise in the case of experimental data, to numerical noise in the case of computer simulations, or in general, to unobserved sources of variability. 
Then, any observation $y_\text{H}$ at an input point $\ve x$ can be expressed in the general form:
\begin{equation}\label{eq:regression_form}
    y_\text{H}(\ve{x}, \varepsilon_\text{H}) = \psi_\text{H}(\ve{x}) + \varepsilon_\text{H}, 
\end{equation}
where $\ve x$ is a realization of $\ve X$, $\psi_\text H(\ve x)$ is an unknown deterministic function, and $\varepsilon_\text H$ is considered to be an additive noise term, independent of $\ve x$ and modeled as a random variable, following some prescribed zero-mean distribution:
\begin{equation}\label{eq:noise_distr_hf}
    \varepsilon_\text H\sim f_{\varepsilon_\text H}(\varepsilon_\text H), \qquad \mathbb{E}\left[\varepsilon_\text H\right] = 0.
\end{equation}
Thus, each noise-contaminated observation in the available HF data set $(\ve\cx_\text{H}, \cy_\text{H})$ can be expressed as:
\begin{equation}\label{eq:hf_noise}
    y_\text{H}(\ve x_\text{H}^{(i)}, \varepsilon_\text{H}^{(i)}) = \psi_\text{H}(\ve{x}_\text{H}^{(i)}) + \varepsilon_\text{H}^{(i)}, \quad i= 1,...\,, N_\text H,
\end{equation}
where $\varepsilon_\text{H}^{(i)}$ is a realization of the noise $\varepsilon_H$.
In the rest of this paper, for the sake of notation conciseness we denote each observation $y_\text{H}(\ve x_\text{H}^{(i)}, \varepsilon_\text{H}^{(i)})$ in the HF data set simply as $y_\text{H}^{(i)}$.

Moreover, let us consider that for the same system there is another information source of lower fidelity, which provides us with the data set $(\ve\cx_\text{L}, \cy_\text{L})$, where
\begin{equation}\label{eq:lf_noise}
    y_\text{L}(\ve x_\text{L}^{(i)}, \varepsilon_\text{L}^{(i)}) = \psi_\text{L}(\ve{x}_\text{L}^{(i)}) + \varepsilon_\text{L}^{(i)}, \quad i= 1,...\,, N_\text L,
\end{equation}
with $\varepsilon_\text L\sim f_{\varepsilon_\text L}(\varepsilon_\text L)$ and $\mathbb{E}\left[\varepsilon_\text L\right] = 0.$
The size $N_\text L$ of this lower-fidelity data set is generally larger than that of the corresponding HF data set. 

A MFSM aims to directly estimate the underlying deterministic HF function $\psi_\text{H}(\ve x)$ with a function $\hat\psi_\text{H}(\ve x)$ by combining all the available variable-fidelity information.
To this end, the LF response can be represented by a classical surrogate model $\hat{\psi}_\text{L}(\ve x)\approx {\psi}_\text{L}(\ve x)$, since the cost of obtaining data from the associated source cannot in general be assumed negligible. 
Then, assuming that the LF model captures the general trend of the underlying HF function $\psi_\text{H}(\ve x)$, we can express the MFSM as a linear combination of the LF surrogate and a discrepancy function $\delta(\ve x)$:
\begin{equation}\label{eq:mf_linear_full}
    \hat{\psi}_\text{H}(\ve x) = \rho(\ve x) \cdot \hat{\psi}_\text{L}(\ve{x}) + \delta(\ve x),
\end{equation}
where $\rho(\ve x)$ is a scaling function. We can simplify \Cref{eq:mf_linear_full} by assuming the scaling function to be a constant $\rho$:
\begin{equation}\label{eq:mf_linear}
    \hat{\psi}_\text{H}(\ve x) = \rho \cdot \hat{\psi}_\text{L}(\ve{x}) + \delta(\ve x).
\end{equation}

As implied in \Cref{eq:mf_linear}, we assume a strict hierarchy of fidelities in our context, meaning that across the entire domain of definition, the higher-fidelity model is more accurate and/or more computationally expensive to evaluate than the lower-fidelity model.
In addition, for the multi-fidelity surrogate model as in \Cref{eq:mf_linear} to outperform a surrogate model built solely from high-fidelity data, it is implicitly assumed that the discrepancy $\delta(\ve x)$ between the HF model and the scaled LF model is less complex, and thus simpler to model, than the HF model itself. 

Our MFSM approach to estimate the noise-free HF function $\psi_\text{H}(\ve x)$ is based on regression; therefore, for the rest of this paper, the terms ``multi-fidelity surrogate model" and ``multi-fidelity regression model" are used interchangeably.
The class of surrogates for the LF model and the discrepancy function $\delta(\ve x)$ can be chosen among a wide range of regression models, including, among others, PCE, GP regression, or neural networks. However, the framework described in this section is \textit{independent} of the particular choice of surrogate modeling methods, and hence, we will refrain from specifying the choice of SM for now.

\citet{Zhang2018} and \citet{Ficini2021} demonstrated that a number of multi-fidelity regression techniques are robust to noise, provided the number of available high-fidelity observations is large enough. 
In other words,
\begin{equation}
    \label{eq:MFReg}
    \lim\limits_{N_H \rightarrow \infty}  \hat \psi_\text H = \psi_\text H .
\end{equation}

When information from more than two levels of fidelity is available, \Cref{eq:mf_linear} can be generalized in a recursive way \citep{kennedy2000}:
\begin{equation}\label{eq:mf_linear_general}
    \hat{\psi}_{s}(\ve x) = \rho_{s} \cdot \hat{\psi}_{s-1}(\ve{x}) + \delta_{s}(\ve x),
\end{equation}
where the predictor of a model at a particular fidelity can be used to construct the predictor for the immediately higher-fidelity model.

\subsection{Combining physical experiments and computer simulations}

Let us consider a set of observations $(\ve\cx_\text{H}, \cy_\text{H})$ obtained through an experimental campaign, each observation of which can be expressed as in \Cref{eq:hf_noise}. 
Moreover, let us assume that the system response can also be predicted by a white-box computational model, such as a system of equations or a complex finite element simulator, denoted as $\cm_\text L(\ve x)$. 

Then, we can assume that experiments can accurately capture the behavior of the system under investigation, but the information they provide is incomplete due to their scarcity.
On the other hand, the available computational model can complement this information by providing a larger amount of data, but at the cost of lower accuracy, due either to inherent model simplifications (e.g. ignoring some physics), or to numerical limitations (e.g. discretization). 
To combine the two, we apply multi-fidelity surrogate modeling, wherein the experimental data is regarded as noise-affected realizations of a high-fidelity model, and the white-box computational model is treated as its low-fidelity equivalent.
Following the rationale described in the previous section, the computational model is replaced by a surrogate model $\widehat\cm_\text L(\ve x)$, constructed with a set of model evaluations $(\ve\cx_\text{L}, \cm_\text L(\ve\cx_\text{L}))$.

Adopting the MF surrogate modeling information fusion scheme introduced previously, a multi-fidelity model predictor aiming to emulate the underlying noise-free HF function output $\psi_\text{H}(\ve x)$ can be expressed as:
\begin{equation}\label{eq:mf_grey}
    \hat{\psi}_\text{H}(\ve x) = \rho \cdot \widehat{\cm}_\text{L}(\ve{x}) + \delta(\ve x).
\end{equation}

\subsection{Confidence and prediction intervals}
Confidence and prediction intervals are powerful means of conveying the uncertainty present in the predictions of a model, thus significantly increasing the informative value of single-point estimates.

If we denote the unobservable error in the MFSM at a given input $\ve x_0$ with respect to the underlying HF model by $e_m(\ve{x}_0)$, then this reads:
\begin{equation}\label{eq:model_err}
    e_\text m (\ve x_0) = \psi_\text H(\ve x_0) - \hat{\psi}_\text{H}(\ve x_0).
\end{equation}
\textit{Confidence intervals} (CIs) express the uncertainty about this model error or, in other words, about where the underlying HF function lies. More precisely, the $(1-2\alpha)$ confidence interval for the underlying HF function at $\ve{x}_0$ is an interval $[\psi_{\text{lo}, \alpha}(\ve{x}_0), \psi_{\text{up}, \alpha}(\ve{x}_0)]$, such that:
\begin{equation}
    \Pro \left [\psi_{\text{lo}, \alpha}(\ve{x}_0) < \psi_\text H (\ve{x}_0) <  \psi_{\text{up}, \alpha}(\ve{x}_0)\right ] = 1-2\alpha,
\end{equation}
where $\alpha$ is typically set equal to $0.05$ for a $90\%$ CI.

This uncertainty is due to the incomplete information provided by the finite-size HF and LF experimental designs, and can be reduced when more data is available. Therefore, it can be interpreted as epistemic uncertainty. 
Assuming that our multi-fidelity regression model is able to accurately represent the underlying HF function in the presence of unlimited data, we have 
$\lim_{N_H \to \infty} (\psi_{\text{up}, \alpha}(\ve{x}_0) - \psi_{\text{lo}, \alpha}(\ve{x}_0)) = 0$, meaning that as $N_\text{H} \to \infty$, the confidence interval for the multi-fidelity regression model predictor collapses to a single value at each point.

Moreover, according to \Cref{eq:regression_form}, a HF output $y_\text H$ at an input $\ve{x}_0$ is expressed as the sum of the underlying HF function $\psi_\text H(\ve{x}_0)$ and a realization of the noise random variable $\varepsilon_\text H$. 
Hence, from \Cref{eq:regression_form,eq:model_err}, the error in the MF model with respect to a noise-contaminated HF observation can be expressed as follows:
\begin{align}
    y_\text H(\ve x_0, \varepsilon_\text{H}) - \hat{\psi}_\text{H}(\ve x_0) &= \psi_\text{H}(\ve{x}_0) + \varepsilon_{\text{H}} - \hat{\psi}_\text{H}(\ve x_0) \nonumber \\
    &= (\psi_\text{H}(\ve{x}_0) - \hat{\psi}_\text{H}(\ve x_0)) + \varepsilon_{\text{H}} \nonumber \\
    &= e_\text m (\ve x_0) + \varepsilon_{\text{H}} . \label{eq:obs_err}
\end{align}
Thus, we can notice that this error is the sum of two independent components: the reducible model error $e_\text m (\ve x_0)$ and the irreducible error $\varepsilon_{\text{H}}$ due to the noise in the HF observations. 
This uncertainty regarding the value of an unseen HF observation is quantified by \textit{prediction intervals} (PIs). 
Similarly to CIs, we can write that the $(1-2\alpha)$ prediction interval for a HF observation at $\ve{x}_0$ is an interval $[y_{\text{lo}, \alpha}(\ve{x}_0, \varepsilon_\text{H}), y_{\text{up}, \alpha}(\ve{x}_0, \varepsilon_\text{H})]$, such that:
\begin{equation}
    \Pro \left [y_{\text{lo}, \alpha}(\ve{x}_0, \varepsilon_\text{H}) < y_\text H (\ve{x}_0, \varepsilon_\text{H}) <  y_{\text{up}, \alpha}(\ve{x}_0, \varepsilon_\text{H})\right ] = 1-2\alpha .
\end{equation}
From \Cref{eq:model_err,eq:obs_err}, it is evident that the PI encloses the corresponding CI, and consequently, the former is wider than the latter. The difference in their width indicates how much we can improve predictions by increasing the amount of training data.

\begin{figure}[h]
    \centering    \includegraphics[width=.6\textwidth, trim = 50 30 50 50]{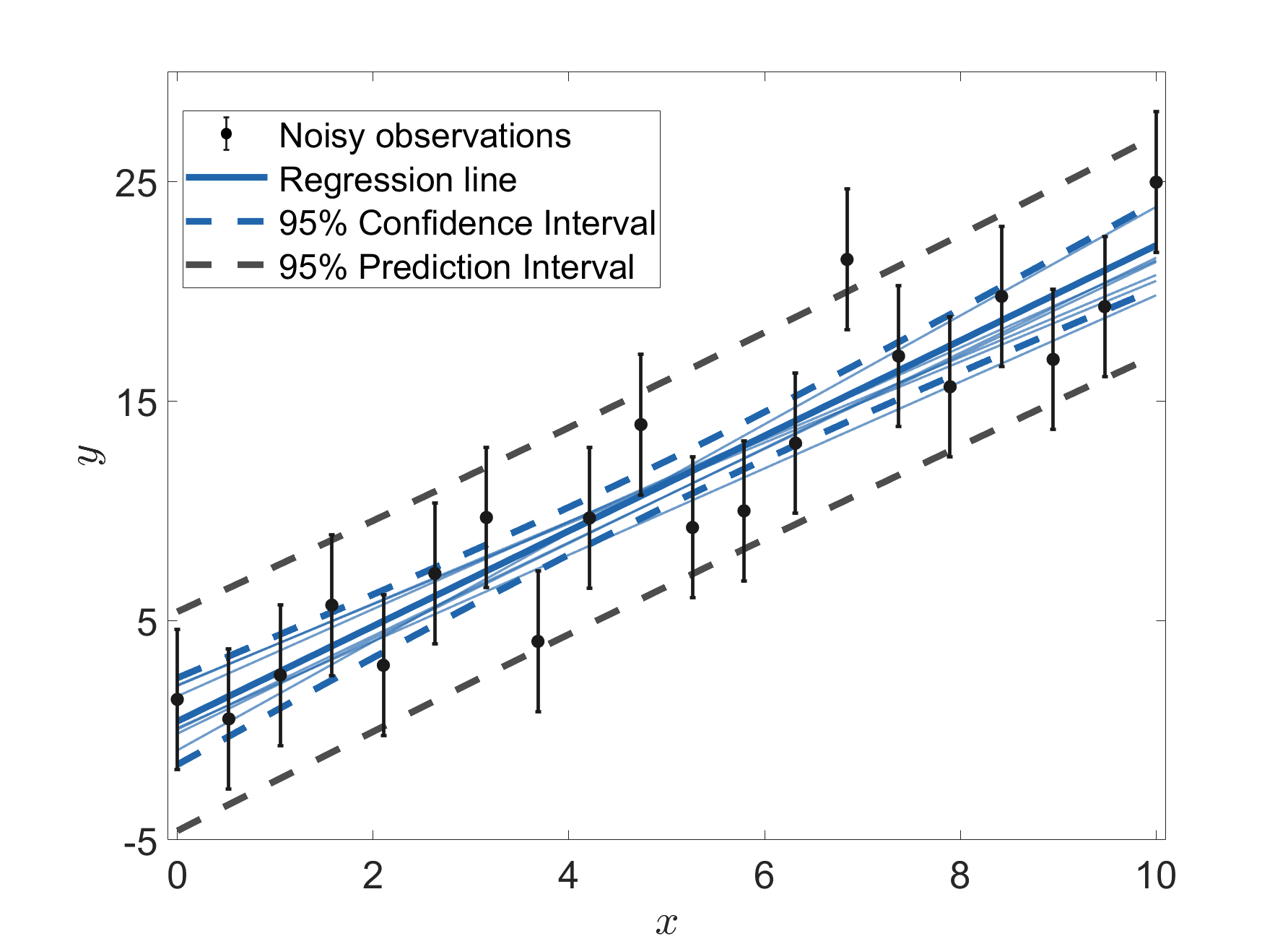}
    \caption{$90\%$ confidence and prediction intervals for the linear regression trained on the illustrated noise-contaminated observations. The thin blue lines represent regression lines for alternative realizations of these observations.}
    \label{fig:CI_PI_sketch}
\end{figure}

\Cref{fig:CI_PI_sketch} illustrates the difference between the CIs and PIs in a single-fidelity linear regression problem with noise-contaminated data. The area between the blue dashed lines is the $90\%$ CI, and shows the uncertainty about where the regression line (blue line) should lie; alternative regression lines trained on different realizations of the same process that generated the current data are represented with thin blue lines. Moreover, the area between the gray dashed lines is the $90\%$ PI, and expresses the uncertainty about where an unseen noise- contaminated observation is likely to fall.

\subsection{Implementation}

\subsubsection{Construction of a multi-fidelity model}
\label{sec:impl_MF_GB}
Starting from a set of high-fidelity experimental data $(\ve \cx_\text H, \cy_\text H)$ and a lower-fidelity computational model $\cm_\text L(\ve x)$, we provide here the methodology to combine the two in a MFSM. 

Our approach uses polynomial chaos expansion as a surrogate model in the hybrid correction scheme introduced in \Cref{eq:mf_grey}, extending the works of \citet{Ng2012,Palar2016,Berchier2016}.
The main motivations behind our choice of PCE as a surrogate in our MFSM methodology include its robustness to noise, its efficiency in terms of training, and its applicability in uncertainty quantification problems.

PCE is a surrogate modeling technique which provides an approximation of a model with finite variance through its spectral representation on a polynomial basis \citep{Xiu2002,Ghanembook2003,LuethenSIAMJUQ2021}.
Let $\ve{X}\in \mathbb{R}^M $ be a random vector with independent components and joint probability density function (PDF) $f_{\ve{X}}(\ve x) = \prod_{i=1}^{M}f_{\ve{X}_i}(\ve x_i),$ where $f_{\ve{X}_i}$ is the marginal PDF of the random variable $\ve{X}_i$.
In practice, the polynomial basis needs to be finite, and the truncated PCE of a computational model $\cm(\ve{x})$ is defined as
\begin{equation}
\label{eq:PCE_trunc}
    \widetilde{\mathcal{M}} \left( \ve{x} \right) = \sum_{\ve\alpha \in \ca} c_{\ve\alpha} {\Psi}_{\ve\alpha} \left( \ve{x} \right),
\end{equation}
where $c_{\ua} \in \mathbb{R}$ are the coefficients of the multivariate polynomials \{$\Psi_{\ua}, \,\ua \in \ca$\}. 
Each polynomial $\Psi_{\ua}$ is the product of univariate polynomials orthogonal with respect to the PDF $f_{\ve{X}_i}$ of the input variable $\ve{X}_i$, and characterized by the multi-index $\ua$.
$\ca \subset \mathbb{N}^M$ is the finite set of multi-indices of the polynomials, and it can be obtained from different truncation schemes, such as total-degree, low-rank or hyperbolic truncation \citep{UQdoc_20_104}. The mean and variance of $\mathcal{M}(\ve X)$ can be approximated by:
\begin{align}    
    \hat\mu_\text{PC} &= c_{\ve 0} \label{eq:mean_PCE}\\
    \hat\sigma^2_\text{PC} &=\sum_{\substack{\ve\alpha \in \ca \\ \ua \neq \ve 0}} c_{\ua}^2. \label{eq:var_PCE}
\end{align}

For the calculation of the PCE coefficients $c_{\ua}$, we adopt a regression-based strategy, as exhaustively reviewed in \citet{LuethenSIAMJUQ2021,LuethenIJUQ2022}, because of its applicability to data-driven problems and robustness to noise \citep{Torre2019}.
Specifically, we opt for a sparse regression solver, least angle regression (LARS) \citep{BlatmanJCP2011}.
Moreover, for the choice of the PCE basis we use degree adaptivity, as well as a total-degree truncation scheme for low-dimensional applications and hyperbolic truncation for higher-dimensional applications.

The first step in constructing our multi-fidelity surrogate model as in \Cref{eq:mf_grey} entails creating a surrogate of the low-fidelity model using PCE. 
For this purpose, we first sample $N_\text L$ realizations $\ve\cx_\text L = \{\ve x_\text{L}^{(1)}, ...,\ve x_\text{L}^{(N_\text L)}\}$ of the input random variables  (e.g. through Latin hypercube sampling; \citet{McKay1979}) and obtain the corresponding model responses $\cy_\text L = \{y_\text{L}^{(1)}, ...,y_\text{L}^{(N_\text L)}\}$. 
We then construct a PCE model 
\begin{equation} \label{eq:lf_pce}
    \widehat{\cm}_\text{L}(\ve{x}) = \sum_{\alpha \in \mathcal{A}_\text{L}} c_{\alpha, \text{L}} \Psi_{\alpha} \left(\ve{x} \right)
\end{equation}
as discussed above.
By using $\widehat{\cm}_\text{L}(\ve{x})$ in place of the original LF model $\cm_\text{L}(\ve{x})$, we eliminate the need for the HF training set to be a subset of the LF one, as from now on we are able to obtain evaluations of $\widehat{\cm}_\text{L}$ at a negligible cost. 
Moreover, we are able to remove any noise that may be present in the LF data in a general MFSM scenario.
However, inaccuracies in the PCE surrogate of the LF model can impact the overall performance of the MFSM.
Accurately approximating the LF model using PCE requires that the LF model response is sufficiently smooth. 
Moreover, highly non-linear LF models may require higher-order polynomials in the PCE to achieve accurate approximations.
The use of degree-adaptive PCE allows for selecting the most appropriate degree, while sparse regression  keeps the number of coefficients to be computed manageable and  minimizes the risk of over-fitting.
Nevertheless, in cases where the cost of evaluating the LF model is negligible, such as when the LF model is given by an analytical function, the LF model ${\cm}_\text{L}(\ve{x})$ can be used directly in the MFSM. This eliminates the need to train $\widehat{\cm}_\text{L}(\ve{x})$, thus avoiding any potential loss of accuracy.

Since our HF experimental data set is given, we can now evaluate $\widehat{\cm}_\text{L}$ at the available corresponding input samples $\ve \cx_\text H = \{\ve x_\text{H}^{(1)}, ...,\ve x_\text{H}^{(N_\text H)}\}$ to obtain 
$\{\widehat{\cm}_\text{L}(\ve x_\text{H}^{(1)}), ...,\widehat{\cm}_\text{L}(\ve x_\text{H}^{(N_\text H)})\}$.
An estimator $\hat\rho$ of $\rho$ in \Cref{eq:mf_grey} can be directly obtained as:
\begin{equation} \label{eq:rho}
    \hat\rho = \mathbb{E}_{\ve x}\left[\frac{y_\text H(\ve x, \varepsilon_\text H)}{\widehat{\cm}_\text{L}(\ve{x})}\right]\approx \frac{1}{N_H}{\sum_{i=1}^{N_H}\frac{y_\text H^{(i)}}{\widehat{\cm}_\text{L}(\ve x_\text{H}^{(i)})}} .
\end{equation}

Moreover, the discrepancy term $\delta(\ve x)$ in \Cref{eq:mf_grey} is given by
\begin{equation}\label{eq:delta}
    \delta(\ve x) = \hat{\psi}_\text{H}(\ve x) - \rho \cdot \widehat{\cm}_\text{L}(\ve{x}).
\end{equation}

Now, using as training data $(\ve{\cx}_\delta, \cy_\delta) = (\ve{\cx}_\text H, \{y_\text H^{(i)}-\hat\rho \widehat{\cm}_\text{L}(\ve x_\text H^{(i)}),\, i=1,..., N_\text H\})$, we train a PCE model $\hat\delta(\ve x)$ for the discrepancy $\delta(\ve x)$, 
\begin{equation} \label{eq:delta_pce}
   \hat\delta(\ve x) = \sum_{\alpha \in \mathcal{A}_\delta} c_{\alpha, \delta} {\Psi}_{\alpha} \left( \ve{x} \right) .
\end{equation}

Lastly, the LF and the discrepancy expansions can be merged into a single expansion, which can be expressed as follows:
\begin{equation} \label{eq:comb_mf_pce}
    \hat{\psi}_\text{H}(\ve x) = \sum_{\alpha \in \mathcal{A}_\text L \cap \mathcal{A}_\delta} (\hat\rho c_{\alpha, \text L} + c_{\alpha, \delta}) {\Psi}_{\alpha} \left( \ve{x} \right) +
    \sum_{\alpha \in \mathcal{A}_\text L \text\textbackslash \mathcal{A}_\delta} \hat\rho c_{\alpha, \text L}  {\Psi}_{\alpha} \left( \ve{x} \right) +
    \sum_{\alpha \in \mathcal{A}_\delta \text\textbackslash \mathcal{A}_\text L} c_{\alpha, \delta}  {\Psi}_{\alpha} \left( \ve{x} \right),
\end{equation}
where $\mathcal{A}_\text L \cap \mathcal{A}_\delta$ is the set of multi-indices present in both the LF and the discrepancy expansions, $\mathcal{A}_\text L \text\textbackslash \mathcal{A}_\delta$ is the set of multi-indices present only in the LF expansion, and $\mathcal{A}_\delta \text\textbackslash \mathcal{A}_\text L$ the set of multi-indices present only in the discrepancy expansion.
This last step is optional, and it is only possible when both $\widehat{\cm}_\text{L}(\ve{x})$ and $\hat\delta(\ve x)$ are PCEs. 
In general, different SMs could be used for either of these two models, in which case the combined expression in \Cref{eq:comb_mf_pce} is not available. Thus, the corresponding step is omitted, and the multi-fidelity predictor is instead expressed as:
\begin{equation}\label{eq:mf_grey_pred}
    \hat{\psi}_\text{H}(\ve x) = \hat\rho \cdot \widehat{\cm}_\text{L}(\ve{x}) + \hat\delta(\ve x).
\end{equation}

According to the categorization by \citet{Peherstorfer2018}, as discussed in the introduction of this paper, this multi-fidelity surrogate modeling method falls under the adaptation category, since the LF data is initially used to construct the LF surrogate model $\hat{\psi}_\text{L}$, which is then corrected through a discrepancy function and a scaling factor using HF model evaluations to predict the HF model.

In summary, the construction of the multi-fidelity surrogate model from a HF experimental data set $(\ve\cx_\text H, \cy_\text H)$ and a LF computational model $\cm_\text L(\ve x)$ according to \Cref{eq:mf_grey} involves the following main steps:
\begin{enumerate}
    \item Use sampling to obtain an ED $(\ve\cx_\text L, \cy_\text L) = (\ve\cx_\text L, \cm_\text L(\ve\cx_\text L))$ for the LF model;
    \item Train a PCE model $\widehat{\cm}_\text{L}(\ve{x})$ on $(\ve\cx_\text L, \cy_\text L)$;
    \item Evaluate $\widehat{\cm}_\text{L}(\ve{x})$ at the available HF parameter sets $\ve\cx_\text H$ to obtain \\
    $\{\widehat{\cm}_\text{L}(\ve x_\text{H}^{(i)}),\, i=1,...,N_\text H\}$;
    \item Estimate $\hat\rho = \mathbb{E}\left[\frac{y_\text H(\ve x, \varepsilon_\text H))}{\widehat{\cm}_\text{L}(\ve x)}\right] \approx \frac{1}{N_H}\sum\limits_{i=1}^{N_H}\frac{y_\text H^{(i)}}{\widehat{\cm}_\text{L}(\ve x_\text{H}^{(i)})}$ ;
    \item Construct a PCE estimator $\hat\delta(\ve x)$ for the discrepancy function, using the ED \\
    $(\ve{\cx}_\text H, \{y_\text H^{(i)}-\hat\rho \widehat{\cm}_\text{L}(\ve x_\text H^{(i)}), i=1,..., N_\text H\})$;
    \item Use the computed $\widehat{\cm}_\text{L}(\ve{x}), \, \hat\rho, \, \hat\delta(\ve x)$ for the MF predictor as in \Cref{eq:mf_grey_pred}. Optionally, merge the LF and discrepancy expansions into one PCE.
\end{enumerate}

In a broader multi-fidelity setting, a low-fidelity data set, or even a pre-trained surrogate of the LF model (not necessarily a PCE model) can be available instead of a LF computational model. In this situation, one can follow the same procedure to construct the MFSM, by simply omitting Step $1$, and in the second case also Step $2$. 
Furthermore,  when a high-fidelity computational model is available instead of a HF data set, an additional step precedes Step~$1$. This consists in using sampling to obtain a HF experimental design $(\ve \cx_\text H, \cy_\text H)$.
These adaptations accommodate the general multi-fidelity case, which need not strictly adhere to the framework of combining physical experiments and computer simulations.

Please note that, in the methodology described above, $\rho$ and $\delta(\ve x)$ are estimated successively in two separate steps. However, in principle, estimating them jointly is also possible. One approach is to first determine the basis functions for the PCE $\hat\delta(\ve x)$, defined by the truncation set $\mathcal{A}_\delta$, by following Steps 1-5. Then, considering $\widehat\cm_\text L (\ve x)$ as another basis function in the expression
\begin{equation}
    \hat{\psi}_\text{H}(\ve x) = \rho \cdot \widehat{\cm}_\text{L}(\ve{x}) +  
    \sum_{\alpha \in \mathcal{A}_\delta} c_{\alpha, \delta, \text{new}}  \boldsymbol{\Psi}_{\alpha} \left( \ve{x} \right),
\end{equation}
one can jointly estimate $\rho$ and the coefficients $c_{\alpha, \delta, \text{new}}$ using OLS. 
Another approach is based on alternating least squares \citep{Chevreuil2015}, where the estimates for $\rho$ and $\delta(\ve x)$ are iteratively refined. The starting value for $\hat\rho$ in this joint optimization can be obtained from the 4th step of the algorithm described above.

If the LF model response is close to zero, the current estimator of $\rho$ given in Step 4 can be problematic. In these cases, one can employ an alternative estimator, computed as the ratio of the standard deviations of the HF and the LF models. These standard deviations can be computed, for example, according to \cref{eq:var_PCE} from PCEs trained on the available HF and LF data sets.
Then,
\begin{equation}
    \hat\rho_\text{alt} = \frac{\hat\sigma_\text H}{\hat\sigma_\text L}.
\end{equation}
While this estimator of $\rho$ provides the additional benefit of being invariant to linear transforms between the HF and LF model responses, it can be sensitive to noise in the data. Therefore, one needs to use it with caution, and only in cases where minor noise is expected.

Finally, an alternative method that avoids these risks is similar to the first variation for joint estimation of $\rho$ and the coefficients $c_{\alpha, \delta}$ of the expansion of $\delta(\ve x)$, but instead of performing a two-step joint estimation of $\rho$ and $c_{\alpha, \delta}$, it combines both estimations in a single step. Specifically, $\widehat{\cm}_\text{L}(\ve{x})$ in \Cref{eq:mf_grey_pred} is treated as a basis function alongside the basis functions ${\Psi}_{\alpha} \left( \ve{x} \right)$ of $\hat{\delta(\ve x)}$. Then, a sparse regression technique, such as LARS or subspace pursuit \citep{Dai2009,Diaz2018} is employed to jointly estimate $\rho$ and $c_{\alpha, \delta}$, eliminating the need to fit each of them separately in an initial step.

In the applications discussed below, we opt for the method outlined in Steps 1-6 due to its simplicity. Notably, its performance closely paralleled that of the variations discussed above.

\subsubsection{Construction of confidence and prediction intervals}
\label{sec:impl_CI_PI}
Our methodology for constructing confidence and prediction intervals is based on bootstrapping. 
The bootstrap estimator is used to determine measures of accuracy for statistical estimates, e.g., standard errors, biases, and confidence intervals, by creating multiple data sets from an original one using random re-sampling with replacement \citep{efron1994introduction}. 
It is based on the notion that a bootstrap sample is drawn from the observed data in a way similar to how the observed data set is drawn from an unknown population probability distribution.
Therefore, inference about a population from an observed data set can be performed by making inference about the latter from the resampled bootstrap data sets.

One of the applications of the bootstrap estimator lies in constructing confidence intervals for regression models \citep{freedman1981bootstrapping}. 
Bootstrap methods have been used to construct confidence and prediction intervals for regression model predictions in a number of single-fidelity studies \citep{Heskes1996,kumar2012bootstrap}.
Moreover, while the application of bootstrap to provide local error estimates to PCE model predictions within a single-fidelity context has been previously studied (see \citet{MarelliSS2018}), its usage in the context of MFSM has not yet been explored.

Our methodology for constructing CIs about the underlying HF function for a MFSM involves two main steps. 
The first step for a CI at an arbitrary given input $\ve x_0$ aims at obtaining $N_\text{B}$ MF bootstrap model evaluations $\hat \psi^{*}_{\text{H}, j}(\ve{x}_0), \, j= 1, ..., N_\text{B}$. 
For this purpose, we need to construct $N_\text{B}$ MFSMs from $N_\text{B}$ MF bootstrap data sets, which we obtain by independently resampling pairs from the HF and the LF experimental designs.
The second step consists in constructing the CI based on the available bootstrap model evaluations. 
To this end, we can apply one of several bootstrap variations, thoroughly described in \citet{efron1994introduction,davison_hinkley_1997,Carpenter2000}. 
We choose the percentile method, due to its simplicity, its range-preserving property (i.e. by construction, the produced intervals always remain within the valid bounds of a system's response, as opposed to other bootstrap methods, e.g., the standard normal method), as well as the satisfactory performance it demonstrates in our setting. 

The $(1-2\alpha)$-quantile CI is obtained from the $\alpha$- and $(1-\alpha)$-quantile of the empirical quantile function of $\hat \psi^{*}_{\text{H}}(\ve{x}_0)$:
\begin{equation}
    \left [\psi_{\text{lo}, \alpha}(\ve{x}_0), \psi_{\text{up}, \alpha}(\ve{x}_0)\right ] = \left[\hat \psi^{*[\alpha]}_{\text{H}}(\ve{x}_0), \hat \psi^{*[1-\alpha]}_{\text{H}}(\ve x_0)\right] .
\end{equation}

More formally, the procedure for constructing a CI at $\ve x_0$ entails the following steps:
\begin{enumerate}
    \item Obtain $N_\text{B}$ MF bootstrap model evaluations $\hat \psi^{*}_{\text{H}, j}(\ve{x}_0), \, j=1, ..., N_\text B$:
    \begin{enumerate}
        \item From the HF ED $(\ve \cx_\text H, \cy_\text H)$, create $N_\text{B}$ HF  bootstrap data sets $(\ve \cx^*_{\text H, j}, \cy^*_{\text H, j})$. Each such data set contains $N_\text H$ pairs $(\ve x^{*(b)}_{\text H, j}, y^{*(b)}_{\text H, j}), \, b = 1, ..., N_\text H$, where \\ $(\ve x^{*(b)}_{\text H, j}, y^{*(b)}_{\text H, j})$ is a random sample from $(\ve{\cx}_\text H, \cy_\text H)$, such that
        \begin{equation}
            P\left[(\ve x^{*(b)}_{\text H, j}, y^{*(b)}_{\text H, j}) = (\ve x^{(i)}_\text H, y^{(i)}_\text H)\right] = \frac{1}{N_\text H}, \,\, \text{for}\,\, i=1, ..., N_\text H \, ;
        \end{equation}

        \item Similarly, from the LF ED $(\ve \cx_\text L, \cy_\text L)$, create $N_\text{B}$ LF bootstrap data sets \\ $(\ve \cx^*_{\text L, j}, \cy^*_{\text L, j})$, each one containing $N_\text L$ elements. If $(\ve x^{*(b)}_{\text L, j}, y^{*(b)}_{\text L,j})$ is an element of the $j$-th LF bootstrap data set, then
        \begin{equation}
            P\left[(\ve x^{*(b)}_{\text L,j}, y^{*(b)}_{\text L, j}) = (\ve x^{(i)}_\text L, y^{(i)}_\text L)\right] = \frac{1}{N_\text L}, \,\, \text{for}\,\, i=1, ..., N_\text L\, ;
        \end{equation}

        \item Match one-to-one the HF and LF bootstrap data sets to construct $N_\text{B}$ bootstrap MFSMs $\hat \psi^{*}_{\text{H}, j}(\ve{x}), \, j=1, ..., N_\text B$ ;

        \item Evaluate the bootstrap MFSMs $\hat \psi^{*}_{\text{H}, j}$ at $\ve x_0$ to obtain $\hat \psi^{*}_{\text{H}, j}(\ve{x}_0), \, j=1, ..., N_\text B$ ;
    \end{enumerate}
    \item Construct the $(1-2\alpha)$-percentile CI based on $\hat \psi^{*}_{\text{H}}(\ve{x}_0)$:
        \newline
        Estimate $\left [\psi_{\text{lo}, \alpha}(\ve{x}_0), \psi_{\text{up}, \alpha}(\ve{x}_0)\right ]$ as $\left[\hat \psi^{*[\alpha]}_{\text{H}}(\ve{x}_0), \hat \psi^{*[1-\alpha]}_{\text{H}}(\ve x_0)\right]$, where $\hat \psi^{*[\alpha]}_{\text{H}}(\ve{x}_0)$ and \\ $\hat \psi^{*[1-\alpha]}_{\text{H}}(\ve x_0)$ are the $\alpha$- and $(1-\alpha)$-empirical quantile of $\hat \psi^{*}_{\text{H}}(\ve{x}_0)$.
\end{enumerate}

Moving now to the construction of prediction intervals about an unseen noise-contaminated observation, we can follow the same procedure used for the confidence intervals about the underlying HF function, with the additional step of accounting for the noise inherent in the observations, as follows from \Cref{eq:obs_err}. 
More precisely, accounting for the noise comprises a two-step process. 
First, we need to infer the distribution of $\varepsilon_\text H$ that characterizes the noise present in the HF data (see \Cref{eq:regression_form}).
We do this by obtaining realizations of this noise and then use classical inference to fit and select among a family of possible parametric univariate distributions.

In practice, we can expect that our MF predictor will exhibit some bias, which we denote as $\beta$. 
Then, we can obtain a realization of $\varepsilon_\text H$ by computing the residual for each HF observation:
\begin{equation}\label{eq:resid}
    r^{(i)} = y_\text{H}^{(i)} - \hat\psi_\text H(\ve{x}_\text H^{(i)}) - \beta^{(i)} .
\end{equation}
Moreover, an estimate for the bias $\beta^{(i)}$ is obtained from bootstrap as follows \citep{efron1994introduction}:
\begin{equation}
    \hat \beta^{(i)} = \Esp{\hat \psi^{*}_{\text{H}}(\ve{x}_\text H^{(i)})} - \hat\psi_\text H(\ve{x}_\text H^{(i)}), 
\end{equation}
where $\Esp{\hat \psi^{*}_{\text{H}}(\ve{x}_\text H^{(i)})}$ is the bootstrap expectation, which can be approximated by the sample average
\begin{equation}
    \mu^*(\ve{x}_\text H^{(i)}) = \frac{1}{N_\text{B}}\sum_{b=1}^{N_\text{B}}\hat \psi^{*}_{\text{H}, b}(\ve{x}_\text H^{(i)}) .
\end{equation}
Please note that the bootstrap estimate of bias does not consider biases arising from potential inaccuracies in our regression model, such as those introduced by the truncation of the PCE basis. 
However, it is capable of detecting other biases resulting from, e.g., the estimation of coefficients through sparse regression techniques. 

Substituting $\mu^*(\ve{x}_\text H^{(i)})$ for $\Esp{\hat \psi^{*}_{\text{H}}(\ve{x}_\text H^{(i)})}$ and $\hat \beta^{(i)}$ for $\beta^{(i)}$ in \Cref{eq:resid}, the residual computed at $\ve{x}_\text H^{(i)}$ can be written as:
\begin{align}
    r^{(i)} &= y_\text{H}^{(i)} - \hat\psi_\text H(\ve{x}_\text H^{(i)}) - \hat\beta^{(i)} \nonumber \\
    &= y_\text{H}^{(i)} - \hat\psi_\text H(\ve{x}_\text H^{(i)}) - (\mu^*(\ve{x}_\text H^{(i)}) - \hat\psi_\text H(\ve{x}_\text H^{(i)})) \nonumber \\
    &= y_\text{H}^{(i)} - \mu^*(\ve{x}_\text H^{(i)}).
\end{align}
This means that a realization of $\varepsilon_\text H$ can be estimated as the difference between a HF observation and the bootstrap mean. Having $N_\text H$ noise samples $r^{(i)}$, we use maximum likelihood estimation (MLE) to infer the parameters of a zero-mean distribution. Within the scope of this work, we consider zero-mean variants of the classical Gaussian, Laplace, and Uniform distributions, but in the general case any zero-mean symmetrical or non-symmetrical distribution could be considered. 
Finally, we use the Bayesian information criterion (BIC; \citet{Schwartz1978}) to select the most appropriate distribution among those considered.

The second step for the PI construction at $\ve x_0$ consists in adding a new noise realization $\hat \varepsilon_{\text{H,}j}$ from the estimated noise to each of the bootstrap model evaluations $\hat\psi^{*}_{\text{H,}j}(\ve{x}_0)$ obtained in Step~1.d of the CI construction process to obtain a new noisy HF realization:
\begin{equation}
    \hat y^{*}_{\text{H}, j}(\ve{x}_0, \hat \varepsilon_{\text{H}}) = 
    \hat \psi^{*}_{\text{H}, j}(\ve{x}_0) +  \hat \varepsilon_{\text{H}, j}, \, \text{for } j = 1,..., N_\text{B}.
\end{equation}
Finally, similarly to the CI, the $(1-2\alpha)$-quantile PI is obtained from the $\alpha$- and $(1-\alpha)$-quantile of the empirical quantile function of $\hat y^{*}_{\text{H}}(\ve{x}_0, \hat \varepsilon_{\text{H}})$:
\begin{equation}
    \left [y_{\text{lo}, \alpha}(\ve{x}_0, \varepsilon_{\text{H}}), y_{\text{up}, \alpha}(\ve{x}_0, \varepsilon_\text{H})\right] 
    = \left[\hat y^{*[\alpha]}_{\text{H}}(\ve{x}_0, \hat \varepsilon_{\text{H}}), \hat y^{*[1-\alpha]}_{\text{H}}(\ve x_0, \hat \varepsilon_{\text{H}})\right] .
\end{equation}

The process for the construction of a PI at $\ve x_0$ is summarized as follows: 
\begin{enumerate}
    \item Obtain $N_\text{B}$ MF bootstrap model evaluations $\hat \psi^{*}_{\text{H}, j}(\ve{x}_0), \, j=1, ..., N_\text B$: \\
    Steps a - d are the same as for the CI construction;
    \item Estimate the irreducible noise $\varepsilon_\text H$ present on the HF data:
    \begin{enumerate}
        \item Obtain $N_\text H$ noise realizations from the residuals: $r^{(i)} = y_\text{H}^{(i)} - \mu^*(\ve{x}_\text H^{(i)}), \\ i=1,...,N_\text H$, where $\mu^*(\ve{x}_\text H^{(i)})$ is the bootstrap mean at $\ve{x}_\text H^{(i)}$;
        \item Infer the noise distribution:
        \begin{enumerate}
            \item Use MLE to fit a zero-mean Gaussian, Laplace, and Uniform distribution to the samples $r^{(i)}$;
            \item Use BIC to choose the most suitable distribution $\hat\varepsilon_\text{H}$;
        \end{enumerate}
    \end{enumerate}
    \item Add a new realization of $\hat \varepsilon_\text H$ to each of the bootstrap model evaluations $\hat\psi^{*}_{\text{H,}j}(\ve{x}_0)$ to obtain new noisy HF realizations:\\
    $\hat y^{*}_{\text{H}, j}(\ve{x}_0, \hat \varepsilon_{\text{H}}) = \hat \psi^{*}_{\text{H}, j}(\ve{x}_0) +  \hat \varepsilon_{\text{H}, j}, \, j = 1,..., N_\text{B}$ ;
    \item Construct the $(1-2\alpha)$-percentile PI based on $\hat y^{*}_{\text{H}}(\ve{x}_0, \hat \varepsilon_{\text{H}})$:
        \newline 
        Estimate $\left [y_{\text{lo}, \alpha}(\ve{x}_0, \varepsilon_{\text{H}}), y_{\text{up}, \alpha}(\ve{x}_0, \varepsilon_{\text{H}})\right ]$ as $\left[\hat y^{*[\alpha]}_{\text{H}}(\ve{x}_0, \hat \varepsilon_{\text{H}}), \hat y^{*[1-\alpha]}_{\text{H}}(\ve x_0, \hat \varepsilon_{\text{H}})\right]$, where \\ $\hat y^{*[\alpha]}_{\text{H}}(\ve{x}_0, \hat \varepsilon_{\text{H}})$ and $\hat y^{*[1-\alpha]}_{\text{H}}(\ve x_0, \hat \varepsilon_{\text{H}})$ are the $\alpha$- and $(1-\alpha)$-quantile of $\hat y^{*}_{\text{H}}(\ve{x}_0, \hat \varepsilon_{\text{H}})$.
\end{enumerate}

\section{Validation and results}
\label{sec:validation}

In this section, the performance of the proposed framework for multi-fidelity surrogate modeling is illustrated on three examples of increasing complexity: an analytical example with one-dimensional input, a case study with a ten-dimensional finite-element model of a truss and its simply supported beam approximation, and a real-world application involving wind turbine simulations. 
In each application, the HF data contain noise, which is either naturally present (wind turbine application) or artificially introduced by us to replicate the real-world noise-contaminated setting (analytical 1-D example and truss model). 
The validation of the proposed framework comprises two parts: assessing the performance of our MFSM and appraising the confidence and prediction intervals, whose construction is described in the previous section.

For the implementation of the PCE models involved in the validation process, we use UQLab \citep{Marelli2014}, a general-purpose uncertainty quantification software implemented in Matlab.

\subsection{MFSM performance and convergence evaluation}
We assess the predictive performance and convergence behaviour of our MFSM using the \textit{normalized validation error}, computed on a test set consisting of $N_\text{test}$ data points that were not used for training, as follows:
\begin{equation}
\epsilon_\text{val} =\frac{\sum_{i=1}^{N_\text{test}} (y_\text t^{(i)}-\hat\psi_\text H(\ve x_\text t^{(i)}))^2}{\sum_{i=1}^{N_\text{test}} (y_\text t^{(i)}-\mu_y)^2} ,
\end{equation}
where $y_\text t^{(i)}$ equals the noise-free HF model response $\psi_\text H(\ve x_\text t^{(i)})$ at the test point \(\ve x_\text t^{(i)}\), when this response is available (analytical 1-D example and truss model), or the noisy HF response $y_\text H(\ve x_\text t^{(i)})$ when the noise-free response is not known (wind turbine application), and \(\mu_y\) is the mean value of the HF response.

The convergence of our MFSM with respect to the HF experimental design size can be investigated by performing simulations with increasing HF ED size, while keeping the LF ED fixed. Due to the statistical uncertainty associated with each HF random design, 50 replications are carried out, considering each time a different independent realization of this experimental design. Box plots are used to provide an aggregated view of the results obtained in all scenarios.

In the subsequent applications, our objective is to assess the added value of the MFSM in comparison to single-fidelity models, and specifically, we aim to determine whether the MFSM exhibits a faster convergence rate. For this purpose, for each HF experimental design and each replication, alongside the MFSM, we construct a PCE surrogate model trained solely the HF data. The same test data is used to compute $\epsilon_\text{val}$, for both the MFSM and the HF PCE model.

\subsection{Performance measures for confidence and prediction intervals}
\label{sec:CI_PI_eval_criteria}
Regarding the evaluation of confidence and prediction intervals, two well-established key indicators are given by the \textit{confidence interval coverage probability} (CICP) and \textit{prediction interval coverage probability} (PICP) respectively, as well as the \textit{average coverage error} (ACE) \citep{Wan2014a}.

If the nominal coverage of a CI is $1-2\alpha$ and the corresponding CI is $\left [\psi_{\text{lo}, \alpha}(\ve{x}), \psi_{\text{up}, \alpha}(\ve{x})\right ]$, one can estimate the CICP associated with this nominal coverage using $\hat C_\alpha$, defined as
\begin{equation}
\label{eq:CICP_estimation}
    \hat C_\alpha = \frac{1}{N_\text{t}} \sum_{i=1}^{N_\text{t}} \mathbbm{1}(\psi_\text{H,t}^{(i)} \in  [\psi_{\text{lo}, \alpha}(\ve{x}_\text{H, t}^{(i)}), \psi_{\text{up}, \alpha}(\ve{x}_\text{H,t}^{(i)}) ]),
\end{equation}
where $(\ve\cx_\text{H,t}, \Psi_\text{H,t})$ is a test set of size $N_\text{t}$ of HF inputs and the corresponding noise-free responses, and $\mathbbm{1}(\cdot)$ is the indicator function, which returns 1 if the condition between parentheses is true, and 0 otherwise. 

To account for the statistical uncertainty associated with the HF and LF random designs, as well as the bootstrap sampling, we perform $N_{rep} = 10$ replications with varying random seed, and compute the mean CICP (MCICP) over these replications:
\begin{equation}
\label{eq:mcicp}
    \bar C_\alpha = \frac{1}{N_\text{rep}} \sum_{j=1}^{N_\text{rep}} \hat C_\alpha^{(j)},
\end{equation}

When the computed confidence intervals are reliable, the MCICP should be close to its nominal value, i.e., $\bar C_\alpha \approx 1-2\alpha$.

The CICP and MCICP can only be estimated when the underlying noise-free HF function is known, thus their computation is generally unfeasible in real-world applications when data contain noise.

Similarly, for a prediction interval with nominal coverage $1-2\alpha$, the mean prediction interval coverage probability (MPICP) can be estimated as follows:
\begin{equation}
\label{eq:mpicp}
    \bar P_\alpha = 
    \frac{1}{N_\text{rep}} \sum_{j=1}^{N_\text{rep}} \hat P_\alpha^{(j)},
\end{equation}
where 
\begin{equation}
\label{eq:PICP_estimation}
    \hat P_\alpha = \frac{1}{N_\text{t}}     \sum_{i=1}^{N_\text{t}} \mathbbm{1}(y_\text{H,t}^{(i)} \in [y_{\text{lo}, \alpha}(\ve{x}_\text{H,t}^{(i)}), y_{\text{up}, \alpha}(\ve{x}_\text{H,t}^{(i)}) ])
\end{equation}
is the estimated PICP for each replication.
Here $(\ve\cx_\text{H,t}, \cy_\text{H,t})$ is a test set of size $N_\text{t}$ of HF inputs and the corresponding noise-contaminated responses. 
Reliable PIs have $\bar P_\alpha \approx 1-2\alpha$.

The ACE metric aims instead to quantify the difference between the actual coverage of an interval and its designated nominal coverage.
The ACE for a CI evaluation is defined as
\begin{equation}
    \text{ACE}_{\text{CI},\alpha} = \text{CICP} - (1-2\alpha).
\end{equation}
Here, we use the MCICP instead of the CICP, and thus, for a CI of nominal coverage $1-2\alpha$ the ACE can be estimated by 
\begin{equation}
\label{eq:ace_ci}
    \hat E_{\text{CI}, \alpha} =  \bar C_\alpha - (1-2\alpha).
\end{equation}

Likewise, the ACE for a PI can be estimated by
\begin{equation}
\label{eq:ace_pi}
    \hat E_{\text{PI}, \alpha} =  \bar P_\alpha - (1-2\alpha).
\end{equation}
An ACE value that is close to zero indicates reliable intervals. Moreover, a positive ACE denotes over-coverage, i.e., the interval actual coverage exceeds its nominal value, whereas a negative ACE indicates under-coverage, i.e., the interval actual coverage is lower than its nominal value. 

\subsection{Analytical 1-D example} \label{sec:appl_1D}
\begin{figure}[h]
    \centering
    \includegraphics[width=.6\textwidth, trim = 100 30 100 30]{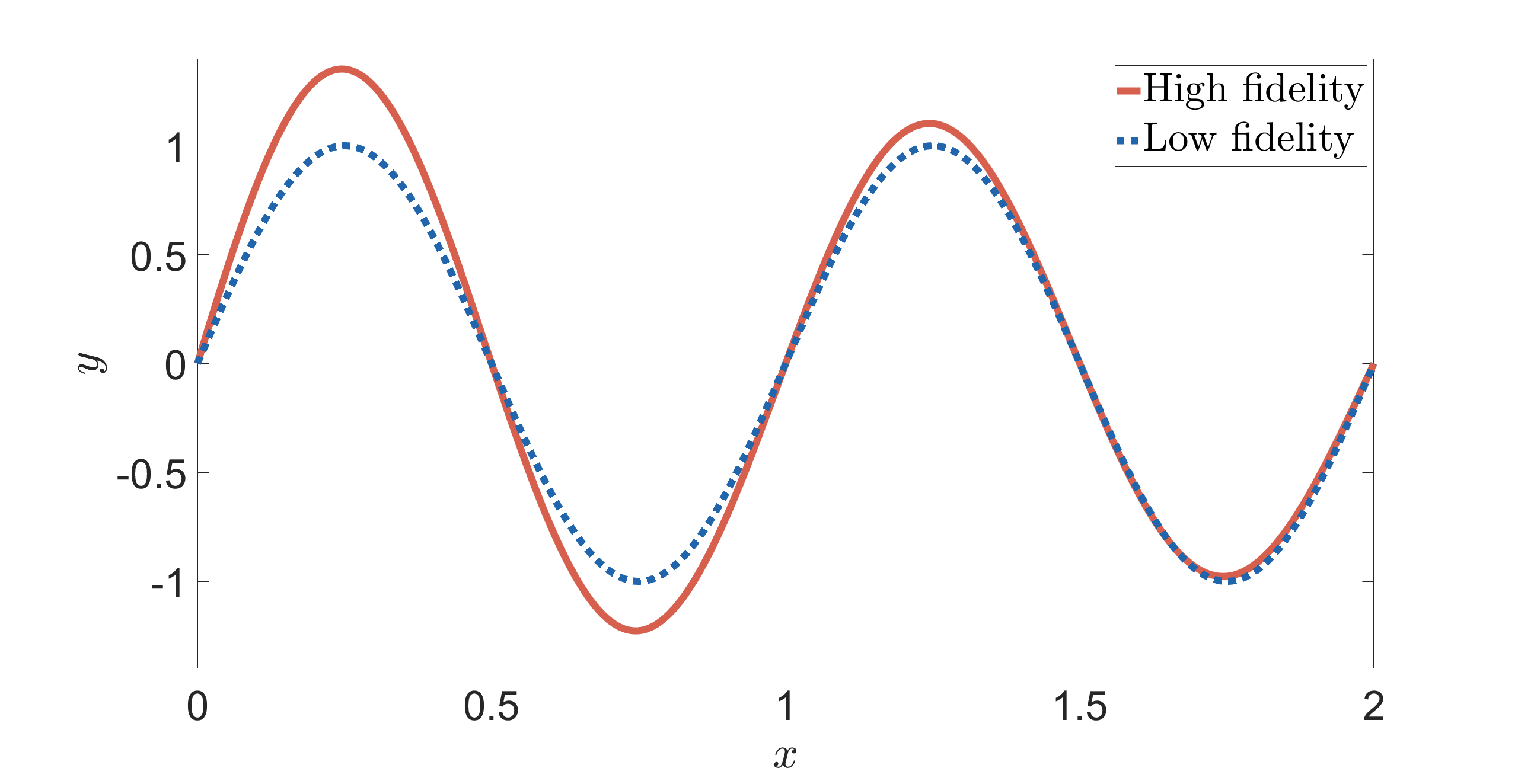}
    \caption{Analytical 1-D example -- Noise-free high- and low-fidelity functions.}
    \label{fig:1d_HF_vs_LF}
\end{figure}

Our first application is an analytical one-dimensional problem which serves well the purpose of visualisation of both the denoising performance and the confidence/prediction interval estimation.
In this application, originally introduced in \citet{Brevault2020a}, the noise-free high- and low-fidelity models are given by:
\begin{align}    
    f_\text{H} &=\left(\frac{x}{4} - \sqrt{2}\right)\sin(2\pi x + \pi) \label{eq:1d_hf_eq}\\
    f_\text{L} &=\sin(2\pi x),
\end{align}
where $ x \sim \mathcal{U}\left[0, 2\right]$. The HF and LF functions are depicted in \Cref{fig:1d_HF_vs_LF}.
For a data set of $1,000$ HF and LF samples, the Pearson correlation coefficient between the HF and LF data is 0.99, while the normalized root mean square error (NRMSE), normalized by the standard deviation of the HF model, is 0.18.

We artificially contaminate the HF data with additive noise that follows a Gaussian distribution $\varepsilon_{\text{H}} \sim \mathcal{N}(0,\,\sigma_{\varepsilon_\text H})$.

\subsubsection{MFSM performance and convergence}
We first assess the performance of the MFSM  under varying levels of noise in the HF data. To this end, we compute the validation error $\epsilon_\text{val}$ for the cases where $\sigma_{\varepsilon_\text H}$ is set to $1\%$, $5\%$, $10\%$, and $20\%$ of the standard deviation $\hat\sigma_\text H$ of the noise-free HF model, obtained from a PCE trained on $1,000$ noise-free HF data points, as described in, e.g., \citet{BlatmanJCP2011}.
As $\hat\sigma_\text H = 0.828$, the numerical values for $\sigma_{\varepsilon_\text H}$ read: (a) $\sigma_{\varepsilon_\text H} = 0.008$, (b) $\sigma_{\varepsilon_\text H} = 0.041$, (c) $\sigma_{\varepsilon_\text H} = 0.083$, (d) $\sigma_{\varepsilon_\text H} = 0.166$, respectively.   

\begin{figure}[h]
     \centering
     \subfigure[$\sigma_{\varepsilon_\text H} = 0.01 \, \hat\sigma_{\text H}$]{
         \includegraphics[width=.475\textwidth, trim = 125 20 10 10]{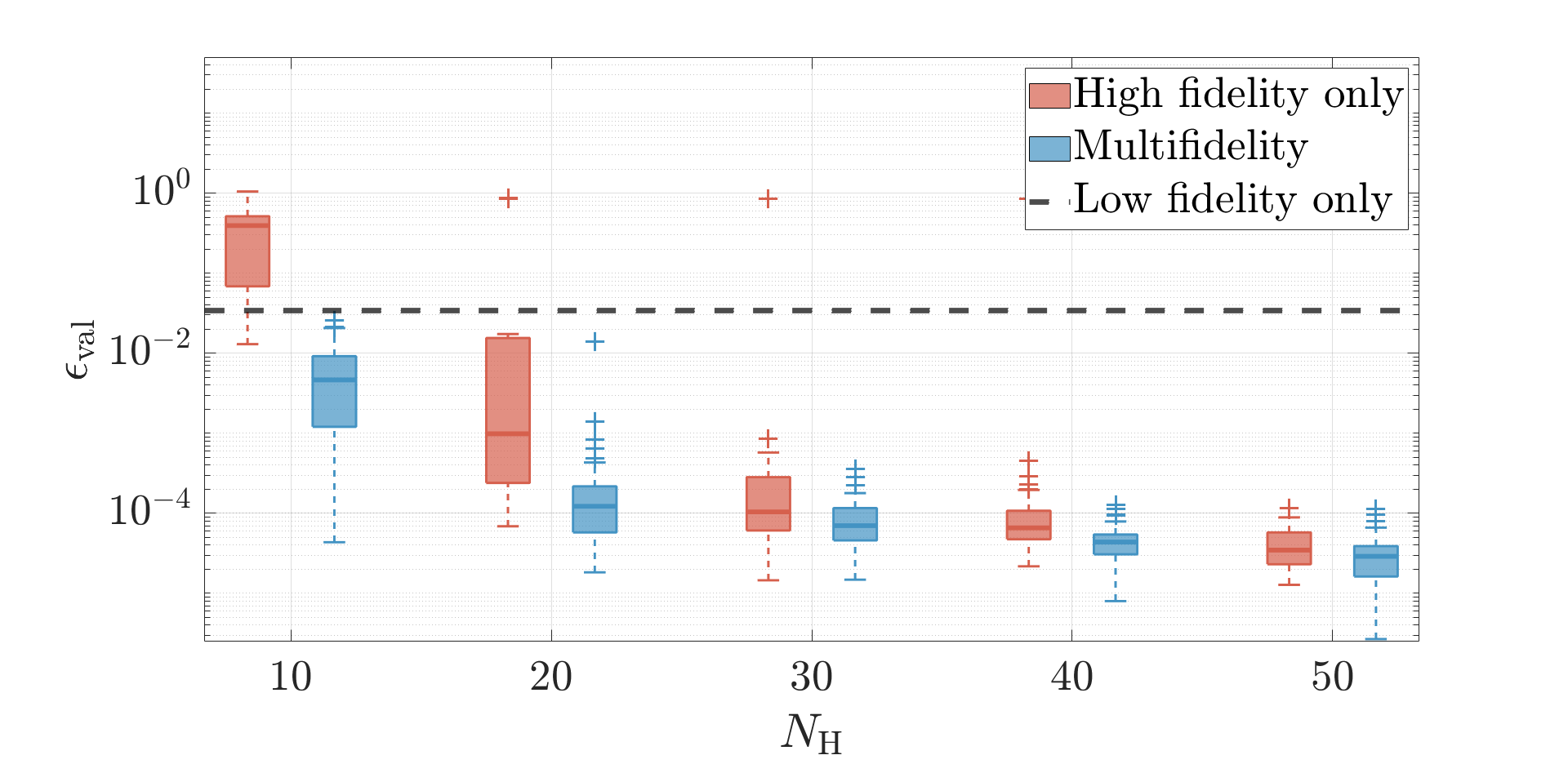}
         \label{fig:noise_1}
     }%
    \subfigure[$\sigma_{\varepsilon_\text H} = 0.05 \, \hat\sigma_{\text H}$]{
         \includegraphics[width=.475\textwidth, trim = 125 20 10 10]{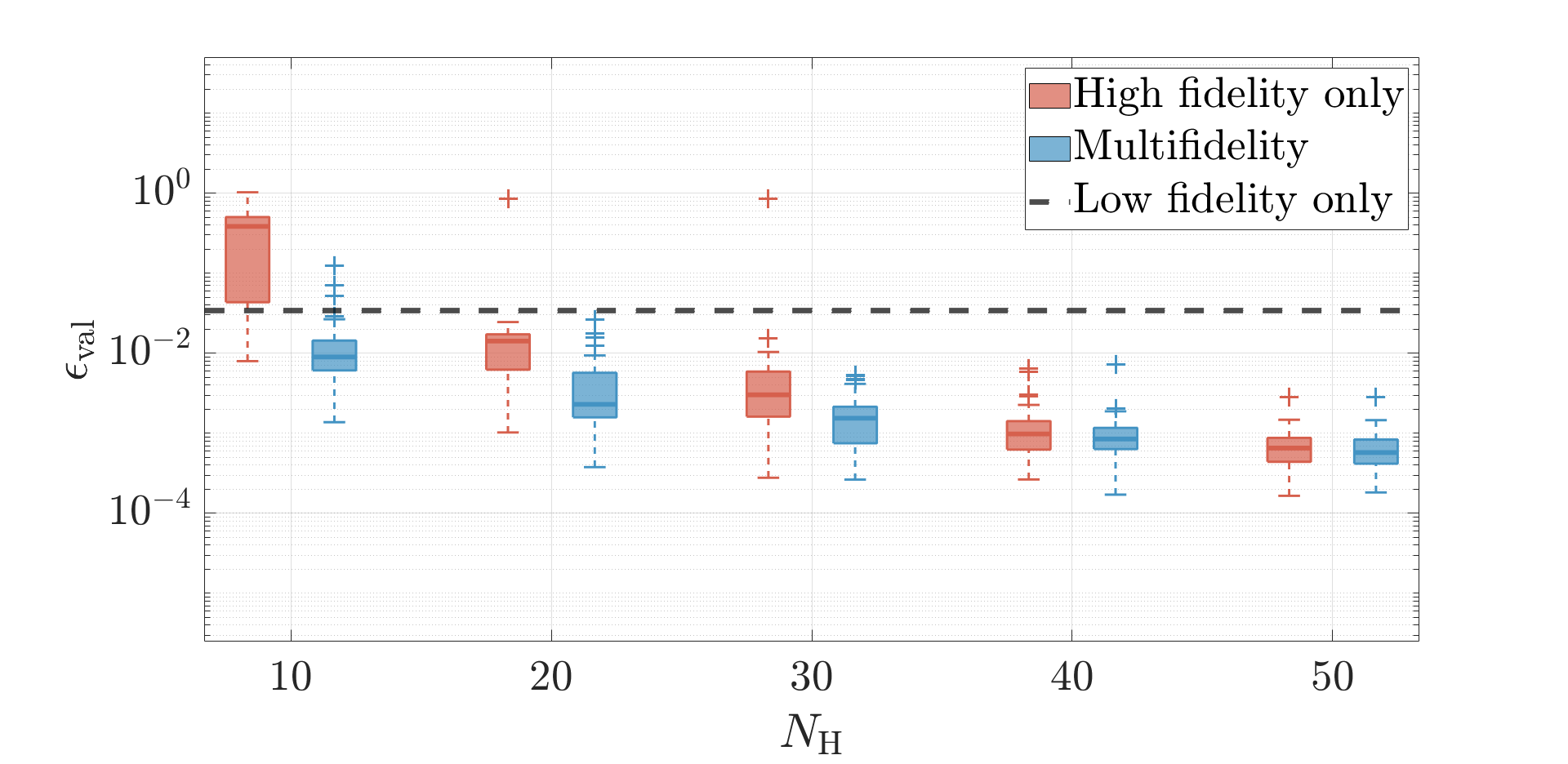}
         \label{fig:noise_2}
    }
    \subfigure[$\sigma_{\varepsilon_\text H} = 0.1 \, \hat\sigma_{\text H}$]{
         \includegraphics[width=0.475\textwidth, trim = 125 20 10 10]{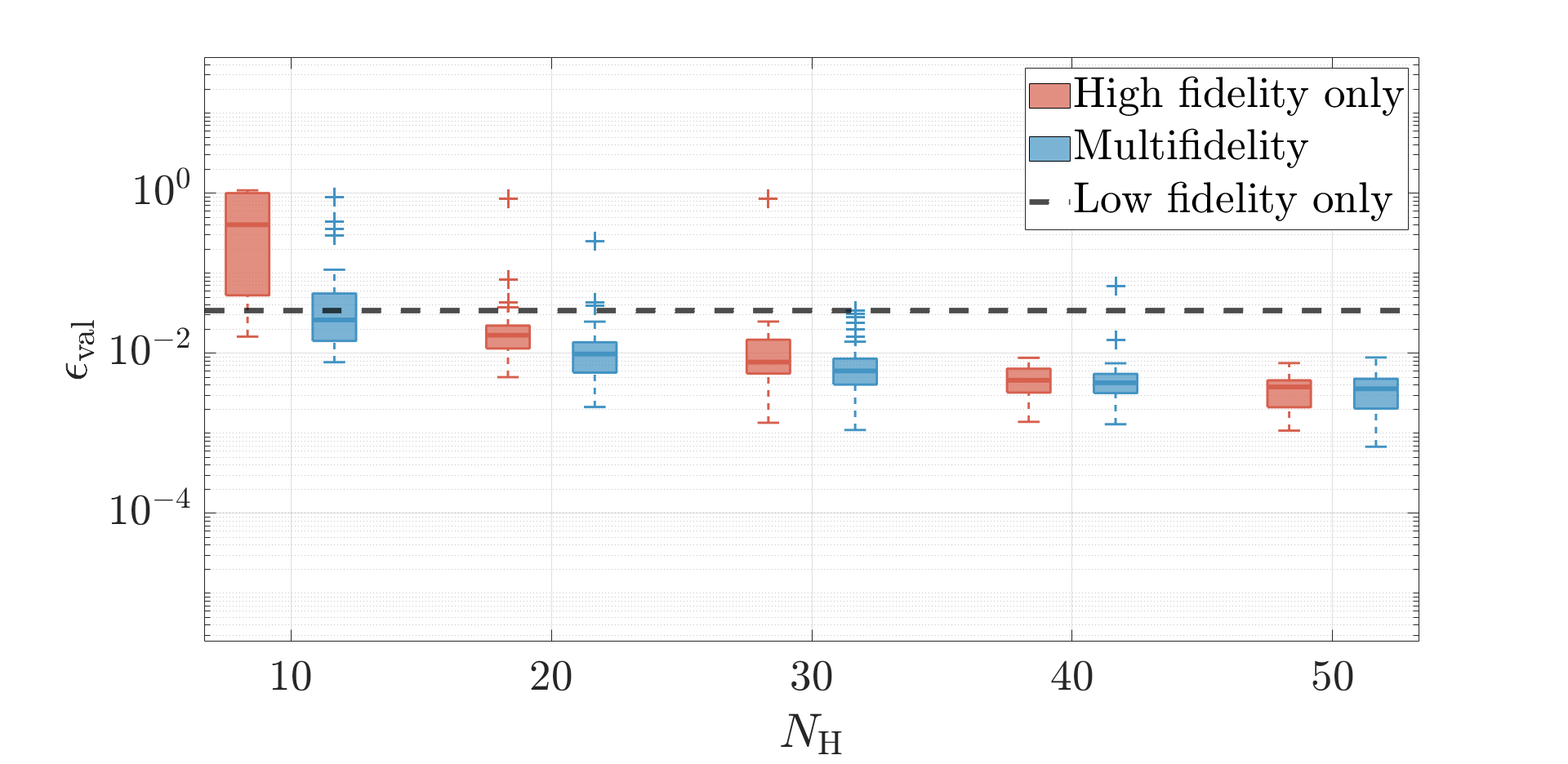}
         \label{fig:noise_3}
    }%
    \subfigure[$\sigma_{\varepsilon_\text H} = 0.2 \, \hat\sigma_{\text H}$]{
         \includegraphics[width=.475\textwidth, trim = 125 20 10 10]{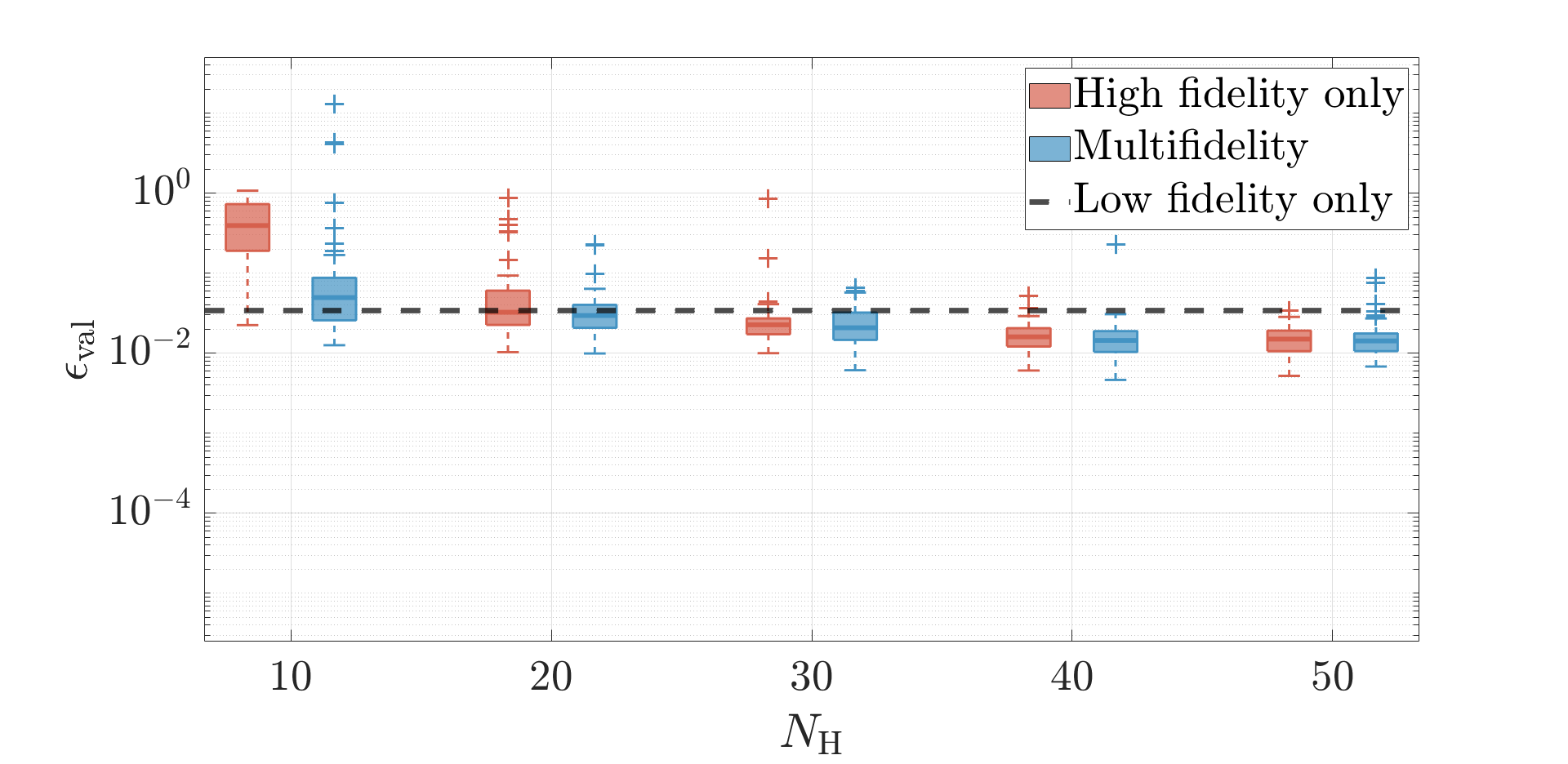}
         \label{fig:noise_4}
    }
        \caption{Analytical 1-D example -- Convergence of the validation error $\epsilon_\text{val}$ for increasing amount of HF training data under varying levels of noise in the HF data, 
        $\varepsilon_\text H \sim \mathcal{N}(0,\, \sigma_{\varepsilon_{\text H}})$. Comparison of our MFSM (blue boxes) with a PCE model trained on HF data only (red). The dashed lines are the corresponding errors of a PCE model trained on LF data only.}
        \label{fig:1-d_boxplot}
\end{figure}

Here, for each noise level, the HF ED varies from 10 to 50 data points, while the LF ED is fixed in all experiments to 100 data points. 
For the choice of basis in all PCEs involved in this example, we use degree adaptivity from degree 1 up to degree 15.
For the computation of the validation errors shown in \Cref{fig:1-d_boxplot}, we use $N_{test} = 10^5$ noise-free HF data points generated using LHS in the input space.

The box plots in \Cref{fig:1-d_boxplot} show the comparison between our MFSM and the PCE surrogate model trained solely on the corresponding HF data available in each case. In addition, the error of the PCE model trained solely on the available LF data is represented by dashed lines and serves as a baseline for comparison. Both single-fidelity PCE models have the same specifications as the PCE employed for the MFSM.

We observe that from as few as 10 HF training data points, our MFSM approach outperforms surrogate models trained solely on either HF or LF data. This distinction becomes particularly evident for lower noise levels within the HF data. 
Hence, when HF data is scarce and data of different fidelities is present, the value of employing MFSMs as opposed to single-fidelity SMs becomes apparent. 
We notice that when sufficient HF data is available, e.g., here, approximately 30 data points for $\varepsilon_\text H \sim \cn(0, \, 0.01\hat\sigma_\text H)$, the MFSM and HF surrogate model performance is similar.
In some instances, when a large amount of HF data is available, the MFSM performance can even be worse than that of the HF surrogate due to the bias introduced by the LF model. As discussed by \citet{Godino2019}, it is not guaranteed that MFSMs always outperform their single-fidelity counterparts.
In addition, as the level of noise in the HF data increases, the difference in performance of the MFSM and the HF SM diminishes.
Indeed, \Cref{fig:noise_4} shows a comparable performance between the two models, regardless of the size of the experimental design.

Moreover, we notice that for all noise levels, the multi-fidelity surrogate model error continuously decreases for increasing $N_\text H$, which indicates the convergence of the MFSM to the underlying noise-free HF model. However, the convergence rate is strongly influenced by the level of noise present in the HF data. As expected, the slowest convergence is observed for the strongest noise (\Cref{fig:noise_4}).

Finally, the scattering of $\epsilon_\text{val}$ of the MFSM, indicated by the box length, is generally smaller for larger HF experimental design sizes. This suggests more stable MFSM models that are less sensitive to the specific choice of the HF experimental design.

\subsubsection{Confidence and prediction intervals}
\label{sec:1d_CIs_PIs}

We now investigate the behaviour and performance of the confidence and prediction intervals for different HF experimental design sizes and different levels of noise in the HF observations. In the following, we use $N_\text B = 1,000$ bootstrap replications to construct the CIs and PIs \citep{Dubreuil2014}.

\begin{figure}[h]
     \centering
     \subfigure[$\sigma_{\varepsilon_\text H} = 0.1 \, \hat\sigma_{\text H}, \: N_\text H = 25$]{
         \includegraphics[width=.475\textwidth, trim = 140 20 90 20]{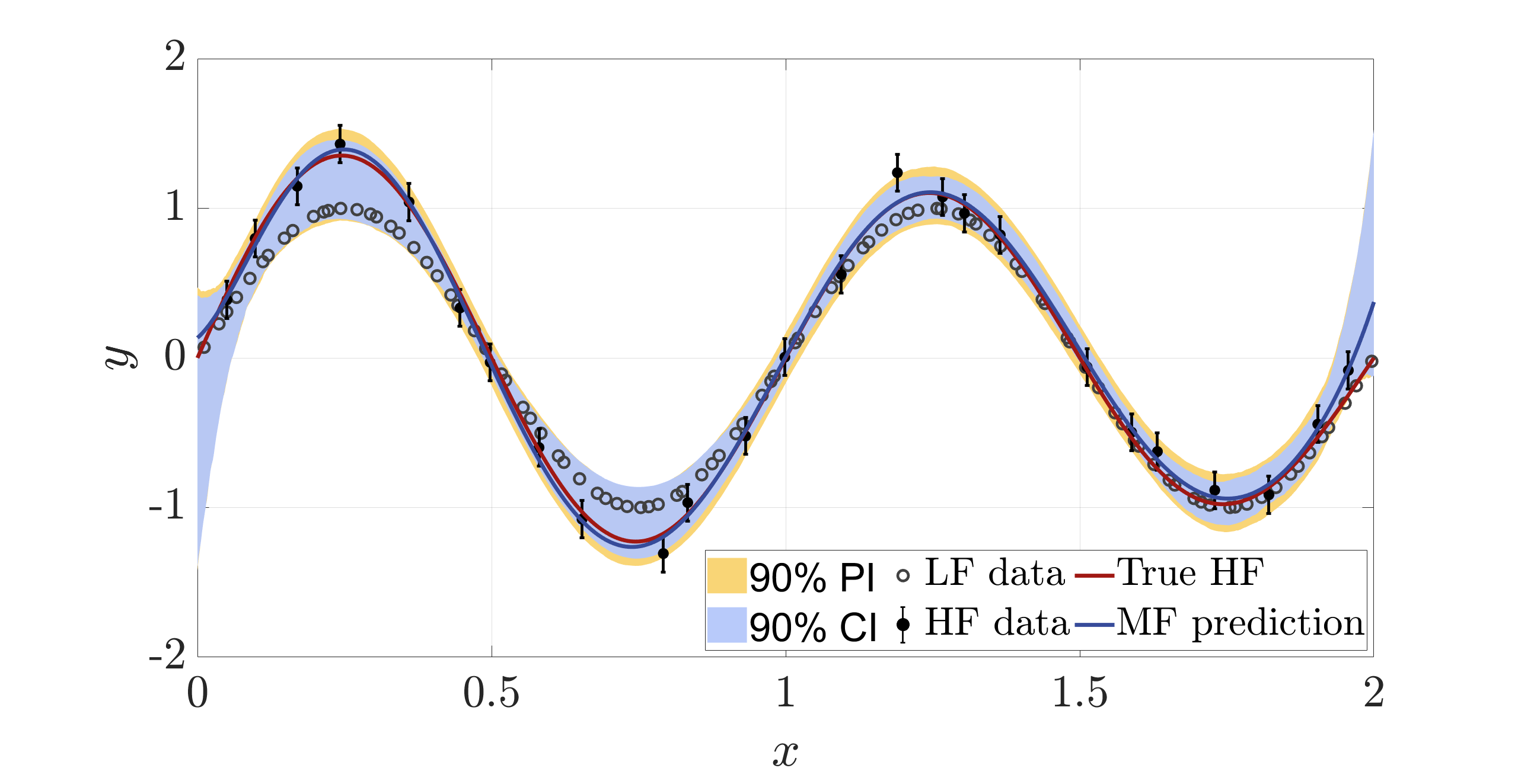}
         \label{fig:1d_nh25_n01}
     }%
    \subfigure[$\sigma_{\varepsilon_\text H} = 0.2 \, \hat\sigma_{\text H}, \: N_\text H = 25$]{
         \includegraphics[width=.475\textwidth, trim = 140 20 90 20]{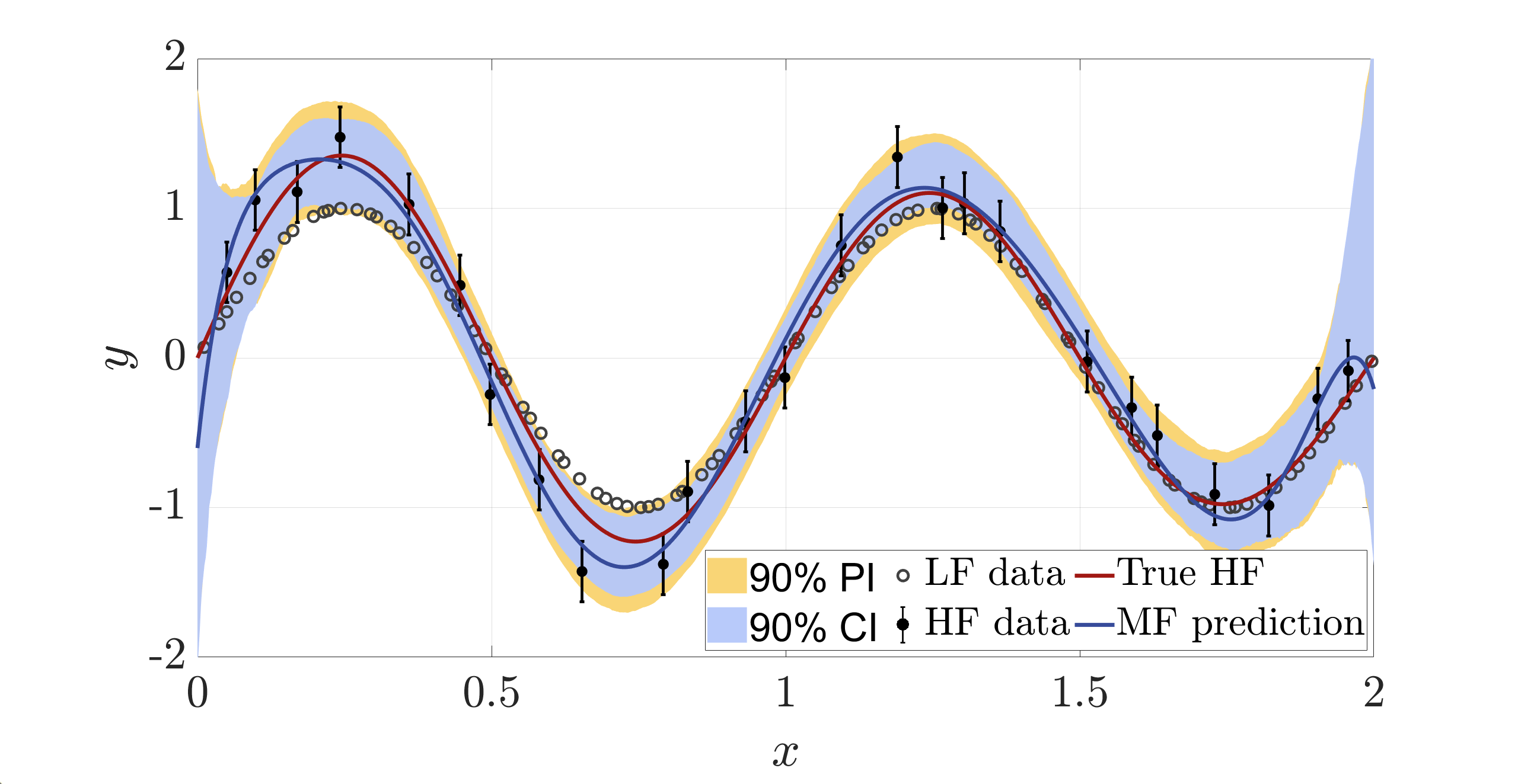}
         \label{fig:1d_nh25_n02}
    }
    \subfigure[$\sigma_{\varepsilon_\text H} = 0.1 \, \hat\sigma_{\text H}, \: N_\text H = 50$]{
         \includegraphics[width=0.475\textwidth, trim = 140 20 90 20]{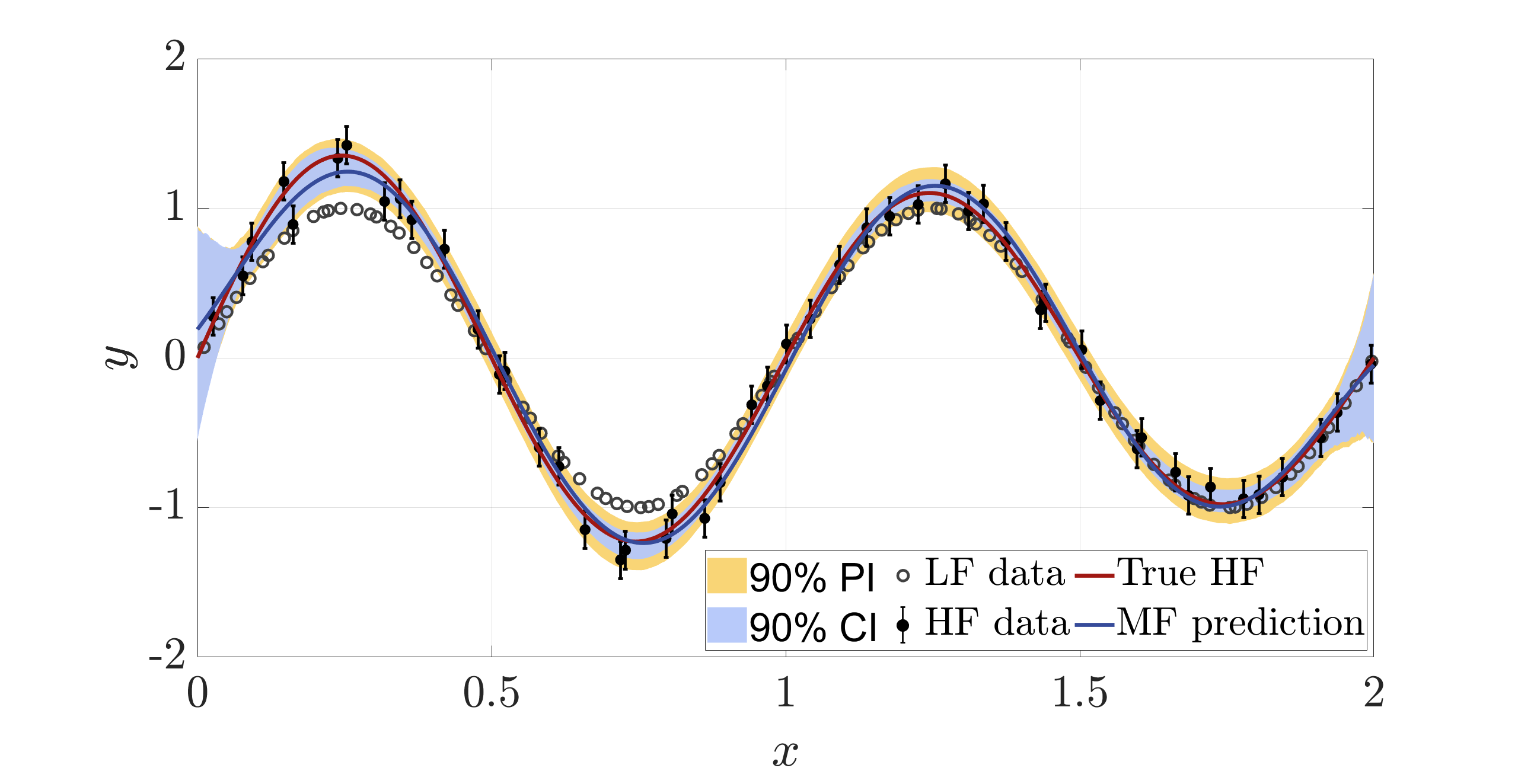}
         \label{fig:1d_nh50_n01}
    }%
    \subfigure[$\sigma_{\varepsilon_\text H} = 0.2 \, \hat\sigma_{\text H}, \: N_\text H = 50$]{
         \includegraphics[width=.475\textwidth, trim = 140 20 90 20]{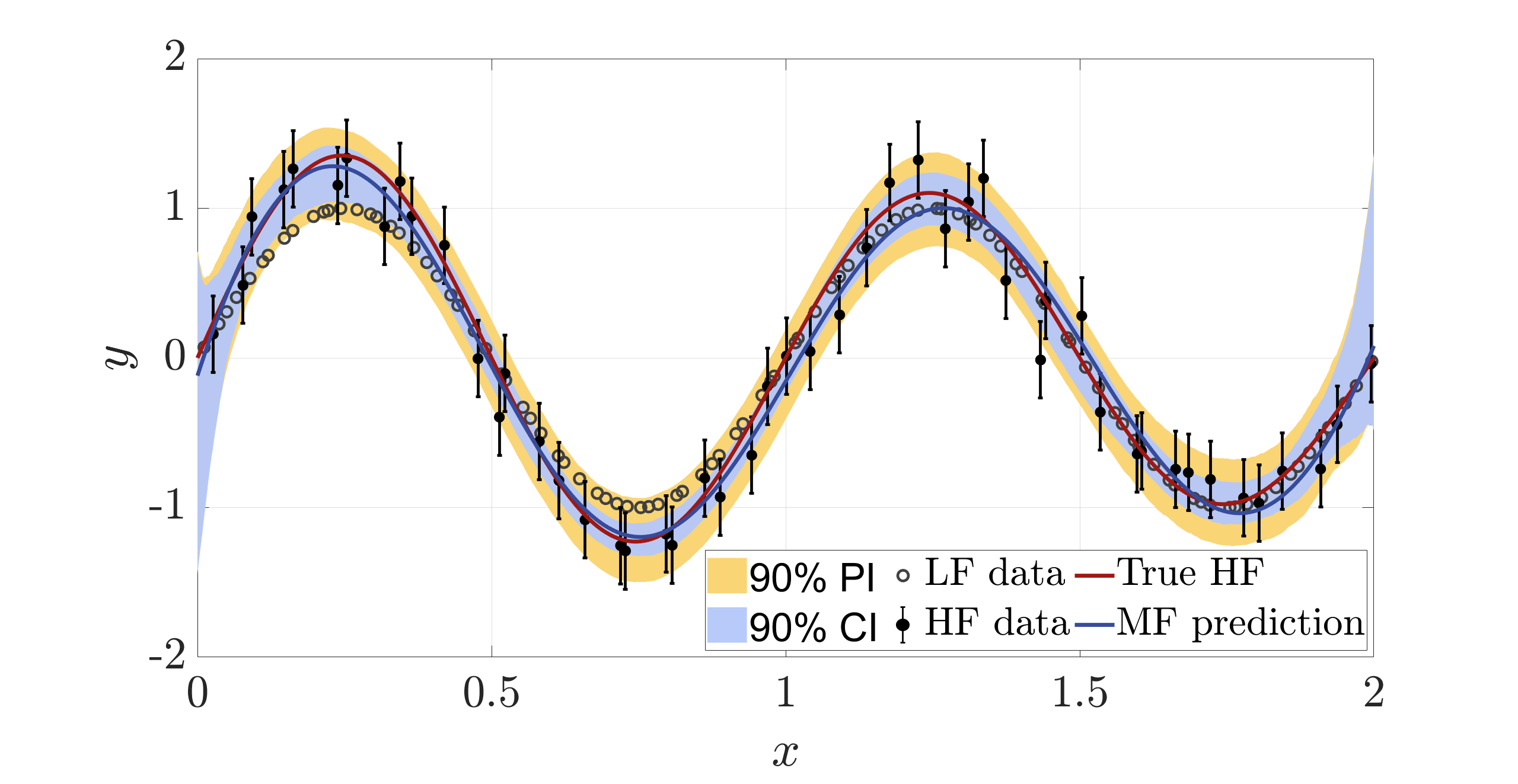}
         \label{fig:1d_nh50_n02}
    }
        \caption{Analytical 1-D example -- $90\%$ confidence and prediction intervals for the MFSMs trained on the illustrated HF and LF data sets. Plots in the same column exhibit the same noise level on the HF data, while the HF ED size increases. Plots in the same row use the same HF ED size and increasing level of noise on the HF data.}
        \label{fig:1d_CIs_PIs}
\end{figure}

\Cref{fig:1d_CIs_PIs} shows the $90\%$ CIs (blue area) and $90\%$ PIs (yellow area) for the MFSM prediction (blue line) in four different cases occurring from all combinations among a realization of higher/lower noise in the HF observations and a realization of a larger/smaller HF ED. Here, again the LF ED is fixed to 100 samples.
The HF and LF training data in each case is visualised by the black error bars and the gray circles, respectively. The error bars depict the $0.9$-quantile of the estimated HF observation noise distribution.
The plots appearing in the same row show that, for increasing noise and same EDs, both the CIs and the PIs become wider. This indicates that both the uncertainty about the underlying HF model and an unseen noise-contaminated HF observation increase.
Moreover, the plots in the same columns reveal that, when the noise remains the same but the HF training data increases, the width of the CIs decreases. This means that our MFSM becomes more confident about where the noise-free HF model lies. 
Also, the uncertainty about an unseen noise-contaminated HF observation, as shown by the PIs, decreases marginally. 
In this case, the PIs and CIs become more distinct, meaning that the uncertainty about a noise-contaminated HF observation is not anymore dominated by the regression model uncertainty.

Let us now proceed to the evaluation of the confidence and prediction intervals that our framework produces.
\Cref{tab:1d_CI_PI_eval} shows the detailed evaluation results including the MCICP and $\text{ACE}_\text{CI}$ for the CI evaluation, as well as the MPICP and $\text{ACE}_\text{PI}$, used for the PI evaluation. 
The MCICP and the $\text{ACE}_\text{CI}$ are estimated as in \Cref{eq:mcicp} and \Cref{eq:ace_ci} respectively, where $N_\text{rep} = 10$ replications with different seeds are performed. 
In each replication, the $\hat C_\alpha^{(j)}$ is computed as in \Cref{eq:CICP_estimation}, using a test set consisting of $N_\text t = 10,000$ data points $\psi^{(i)}_{\text{H,t}}$ from the noise-free HF function, given by \Cref{eq:1d_hf_eq}.
Similarly, to estimate the MPICP and the $\text{ACE}_\text{PI}$, in each of the $N_\text{rep}$ replications, $\hat P_\alpha^{(j)}$ is computed as in \Cref{eq:PICP_estimation}, using a test set with $N_\text t = 10,000$ data points $y^{(i)}_{\text{H,t}}$, where
\begin{equation}
    y^{(i)}_{\text{H,t}} = \psi^{(i)}_{\text{H,t}} + \varepsilon_\text H ^{(i)}.
\end{equation} 
Here $\varepsilon_\text H ^{(i)}$ is a realization of the prescribed noise distribution, $\varepsilon_\text H \sim \cn(0, \sigma_{\varepsilon_\text H})$ with  $\sigma_{\varepsilon_\text H}$ displayed in the second column of the table.

\begin{table}[h]
\centering
    \caption{Analytical 1-D example -- Confidence and prediction intervals evaluation}
    \begin{tabular}{@{}lcccccc@{}}
        \toprule
        $1-2\alpha\:$  & $\:\sigma_{\varepsilon_\text H} \:$ & $\quad N_\text{H}\quad$ & MCICP &  $\text{ACE}_\text{CI}$ & MPICP &  $\text{ACE}_\text{PI}$ \\ 
        \midrule
        \multirow{4}{*}[-1pt]{$0.1$} & $0.1 \,  \hat\sigma_\text H$ & $25$ & $0.118$ & $0.018$ & $ 0.099$ & $-0.001$ \\ 
        & $0.1 \,  \hat\sigma_\text H$ & $50$ & $0.096$ & $-0.004$ & $0.110$ & $0.010$ \\ 
        & $0.2 \,  \hat\sigma_\text H$ & $25$ & $0.125$ & $0.025$ & $0.104$ & $0.004$ \\ 
        & $0.2 \,  \hat\sigma_\text H$ & $50$ & $0.132$ & $0.032$ & $0.104$ & $0.004$ \\
        \midrule
        \multirow{4}{*}[-1pt]{$0.5$} & $0.1 \,  \hat\sigma_\text H$ & $25$ & $0.548$ & $0.048$ & $0.494$ & $-0.006$ \\
        & $0.1 \,  \hat\sigma_\text H$ & $50$ & $0.544$ & $0.044$ & $0.530$ & $0.030$ \\
        & $0.2 \,  \hat\sigma_\text H$ & $25$ & $0.569$ & $0.069$ & $0.509$ & $0.009$ \\ 
        & $0.2 \,  \hat\sigma_\text H$ & $50$ & $0.561$ & $0.061$ & $0.526$ &  $0.026$\\ 
        \midrule
        \multirow{4}{*}[-1pt]{$0.9$} & $0.1 \,  \hat\sigma_\text H$ & $25$ & $ 0.929$ & $0.029$ & $0.888$ & $-0.012$ \\
        & $0.1 \,  \hat\sigma_\text H$ & $50$ & $0.992$ & $0.092$ & $0.905$ & $0.005$ \\
        & $0.2 \,  \hat\sigma_\text H$ & $25$ & $0.960$ & $0.060$ & $0.906$ & $0.006$ \\ 
        & $0.2 \,  \hat\sigma_\text H$ & $50$ & $0.991$ & $0.091$ & $0.919$ & $0.019$ \\ 
        \midrule
        \multirow{4}{*}[-1pt]{$0.95$} & $0.1 \,  \hat\sigma_\text H$ & $25$ & $0.988$ & $0.038$ & $0.941$ & $-0.009$ \\
        & $0.1 \,  \hat\sigma_\text H$ & $50$ & $1$ & $0.050$ & $0.950$ & $0$ \\
        & $0.2 \,  \hat\sigma_\text H$ & $25$ & $1$ & $0.050$ & $0.952$ & $0.002$ \\ 
        & $0.2 \,  \hat\sigma_\text H$ & $50$ & $1$ & $0.050$ & $0.962$ & $0.012$ \\ 
        \bottomrule
    \end{tabular}
    \label{tab:1d_CI_PI_eval}
\end{table}

We can notice that, for all nominal coverage levels and every combination of $\sigma_{\varepsilon_\text H}$ and $N_\text H$, the coverage of the PIs that our method provides is in excellent agreement with the corresponding nominal coverage. 
Specifically, the absolute value of the PI coverage error $\text{ACE}_\text{PI}$ rarely exceeds $1\%$. Moreover,the coverage of the constructed CIs is satisfactory, with the absolute value of $\text{ACE}_\text{CI}$ being most of the times below $6\%$.
The observed error in the CI coverage is almost exclusively due to over-coverage, indicated by a positive $\text{ACE}_\text{CI}$. We can attribute this to the presence of noise in the HF data. Indeed, we can notice that when the noise level is high ($\sigma_{\varepsilon_\text H} = 0.2\, \hat\sigma_\text H$), the CI coverage error increases consistently compared to instances where the noise is lower.
Overall, the PIs achieve coverage much closer to the nominal level rather than the corresponding CIs. 

\begin{figure}[h]
     \centering
     \subfigure[$N_\text H = 20$]{
         \includegraphics[width=.475\textwidth, trim = 140 20 90 20]{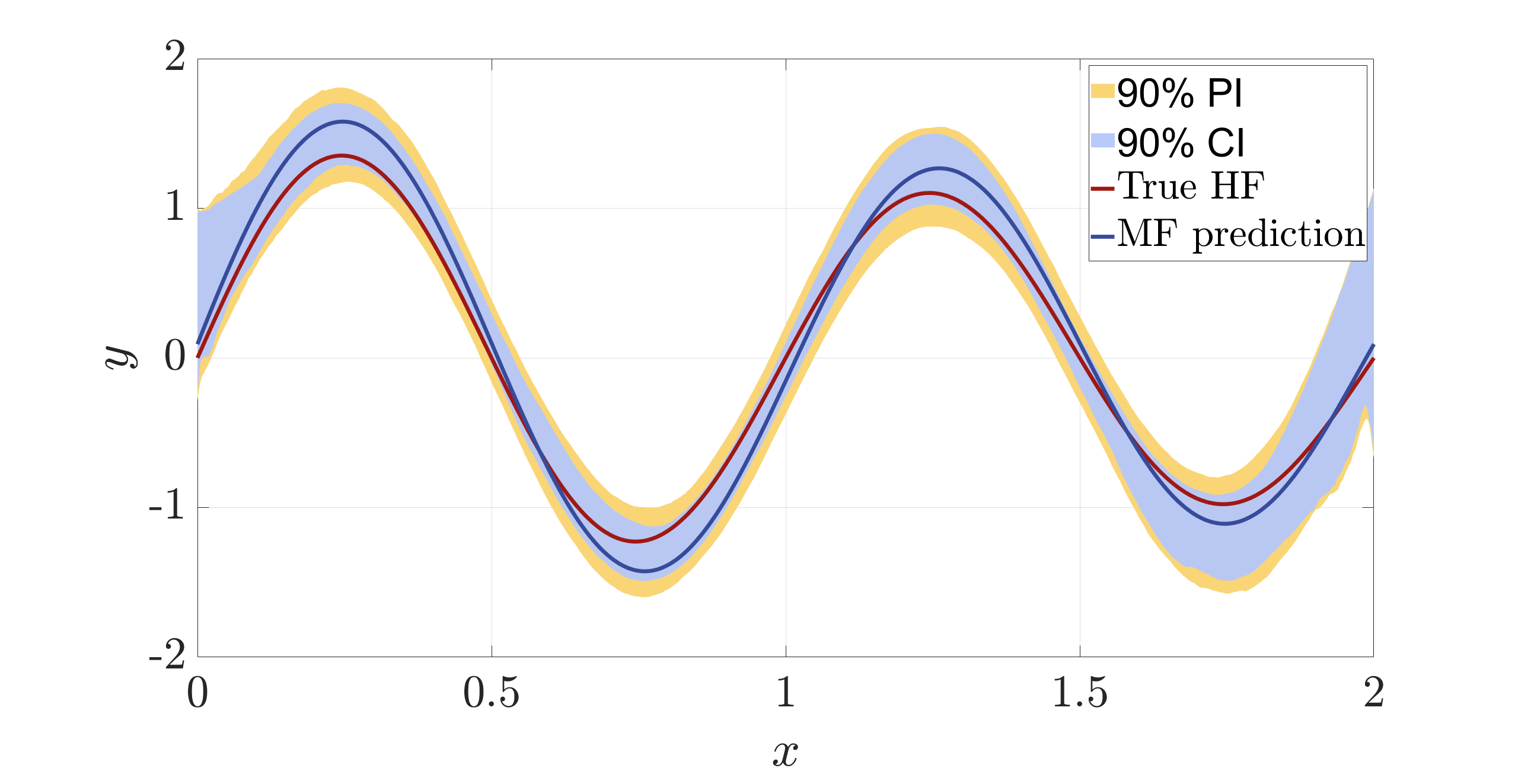}
         \label{fig:1d_as_20}
     }%
    \subfigure[$N_\text H = 100$]{
         \includegraphics[width=.475\textwidth, trim = 140 20 90 20]{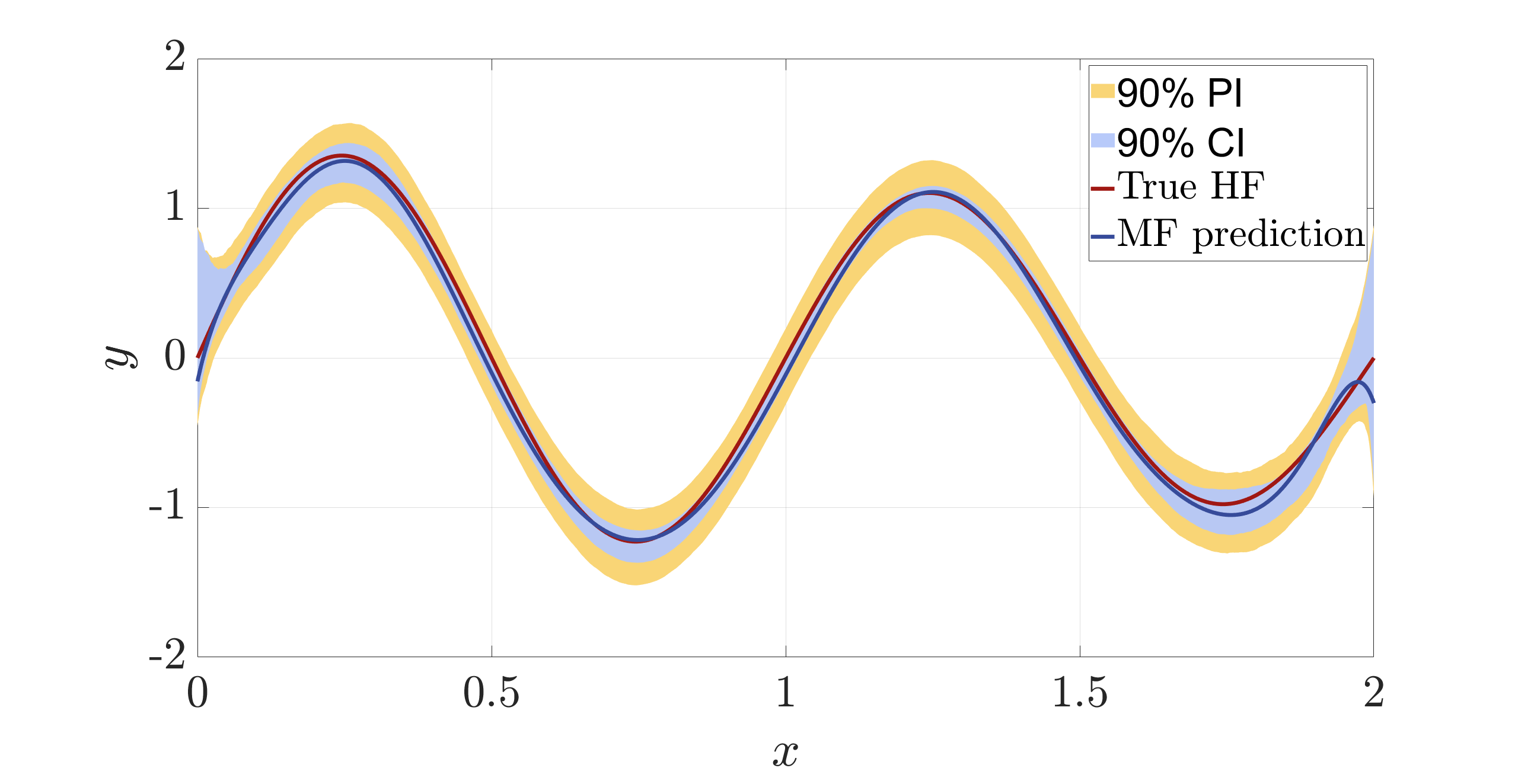}
         \label{fig:1d_as_100}
    }
    \subfigure[$N_\text H = 500$]{
         \includegraphics[width=0.475\textwidth, trim = 140 20 90 20]{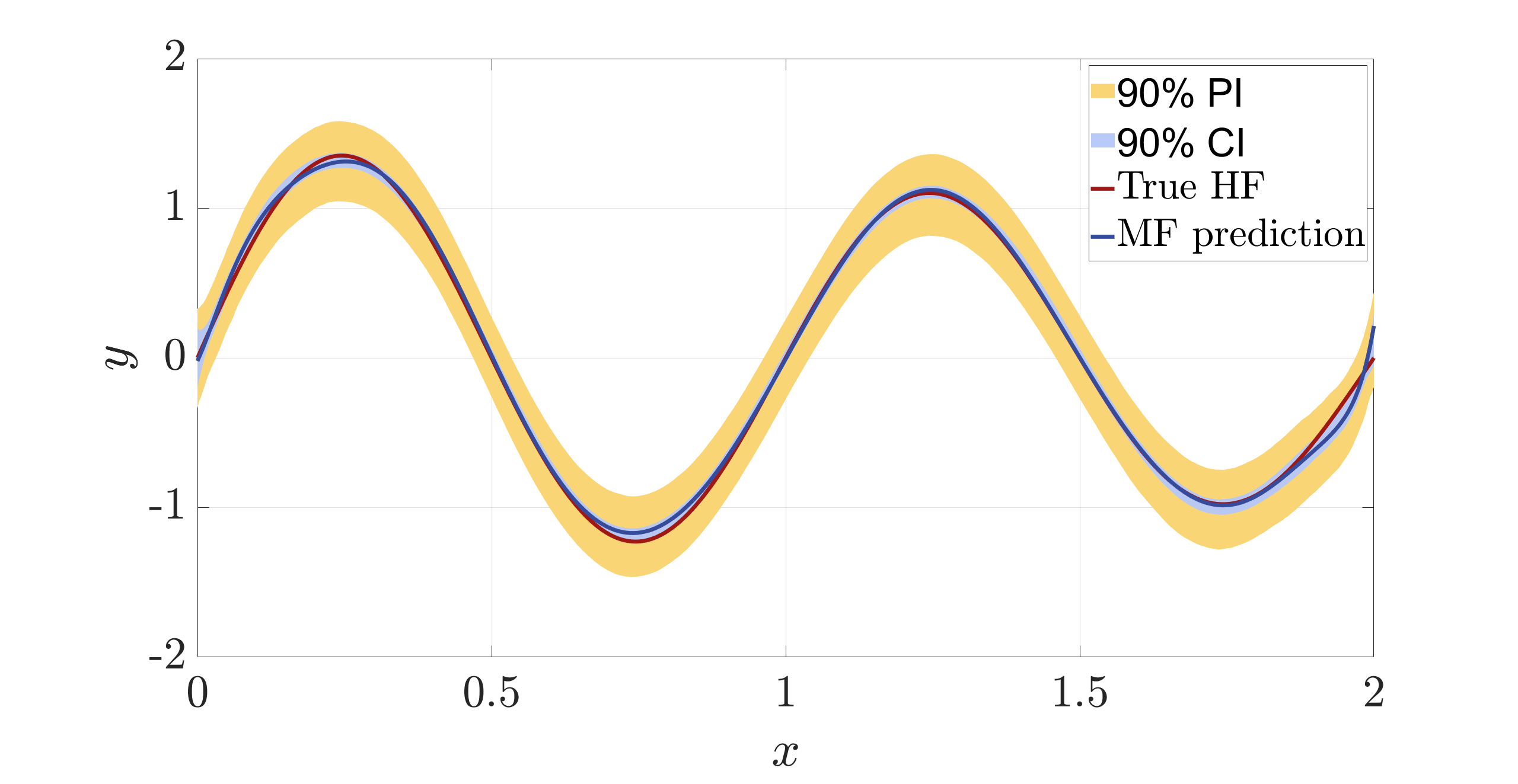}
         \label{fig:1d_as_500}
    }%
    \subfigure[$N_\text H = 2500$]{
         \includegraphics[width=.475\textwidth, trim = 140 20 90 20]{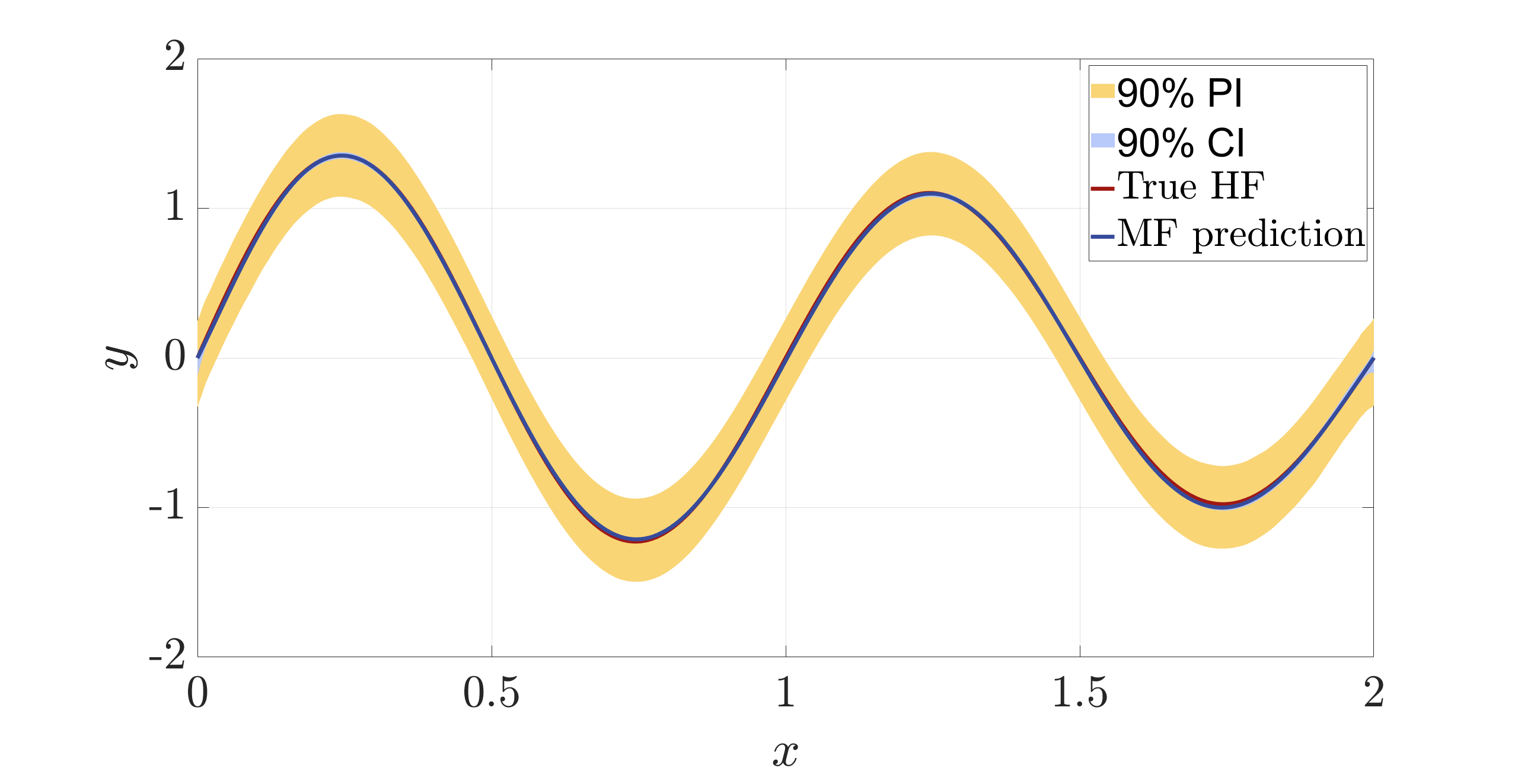}
         \label{fig:1d_as_2500}
    }
        \caption{Analytical 1-D example -- Convergence of the confidence and prediction intervals for increasing HF ED size. In each plot, the blue area corresponds to the $90\%$ CI, the yellow area to the $90\%$ PI, while the red and blue lines depict the true noise-free HF response and the MF prediction respectively. To reduce the visual density of the plot, we did not include the HF and LF training data.}
        \label{fig:1d_CIs_PIs_asymptotic}
\end{figure}

Finally, we investigate the behaviour of the CIs and PIs asymptotically with respect to the HF experimental design size.
\Cref{fig:1d_CIs_PIs_asymptotic} shows the $90\%$ CIs and PIs for four realizations of HF EDs of increasing size, from 20 up to 2500 data points. 
The level of noise is fixed, with $\sigma_{\varepsilon_\text H} = 0.2 \,  \hat\sigma_\text H$, and also fixed is the LF ED to $100$ data points. 
Let us note that despite the LF ED typically being larger than the HF ED in practical applications, the last three out of the four cases do not align with this common scenario. In this study, we intentionally maintain this particular fixed LF ED across all cases to facilitate a focused investigation into the convergence behavior of the CIs and PIs of the MFSM with respect to the HF ED.

We observe that, as the HF ED increases, the CIs tend to converge to the multi-fidelity regression model. 
This behavior aligns with our expectations, and reflects the fact that as more data becomes available, the epistemic uncertainty due to the lack of knowledge decreases and therefore, our MFSM exhibits increased confidence in its predictions. 
As regards the PIs, we notice that they tend to converge to a non-zero width, indicative of the amount of noise in the HF observations. This behavior is also expected, as the noise in the HF observations arises from aleatory uncertainty, and is thus irreducible regardless of the amount of the available training data. Consequently, predictions for unseen observations will inherently carry this uncertainty.

\subsection{Truss model} \label{sec:appl_truss}
\begin{figure}[h]
    \centering    
    \subfigure[HF model: Ideal truss structure.]{
         \centering         \includegraphics[width=0.46\textwidth]{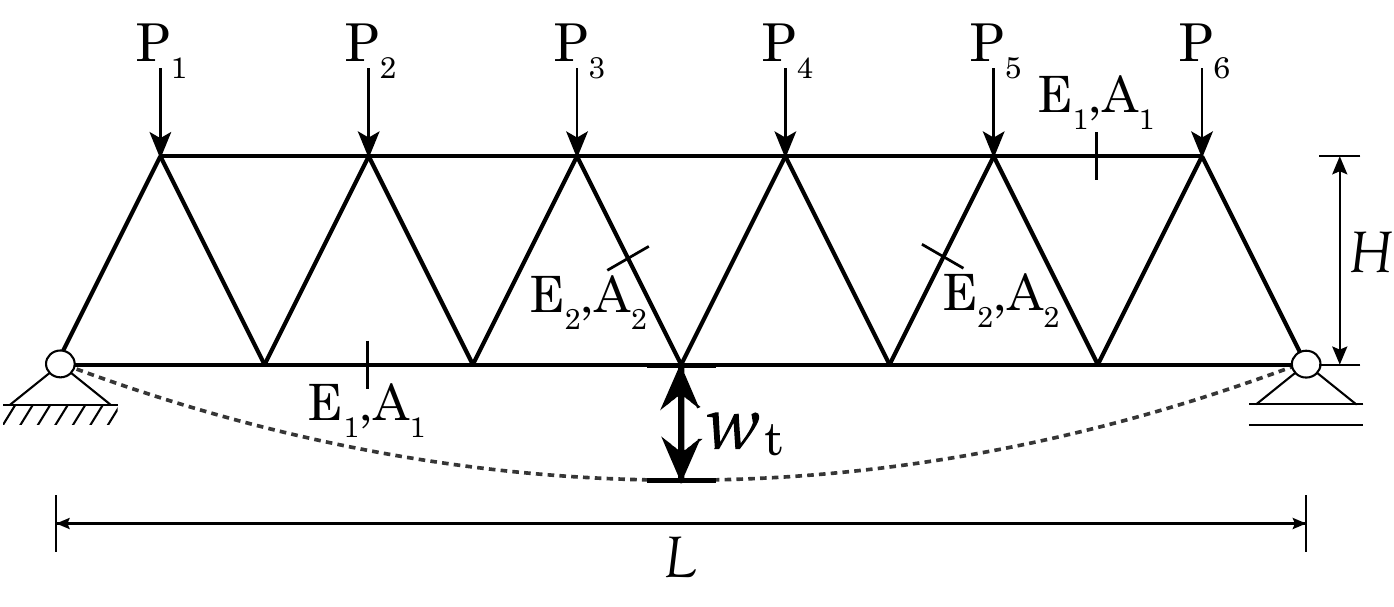}
         \label{fig:truss}
     }
    \hfill
    \subfigure[LF model: Simply supported beam.]{
         \centering       \includegraphics[width=.46\textwidth]{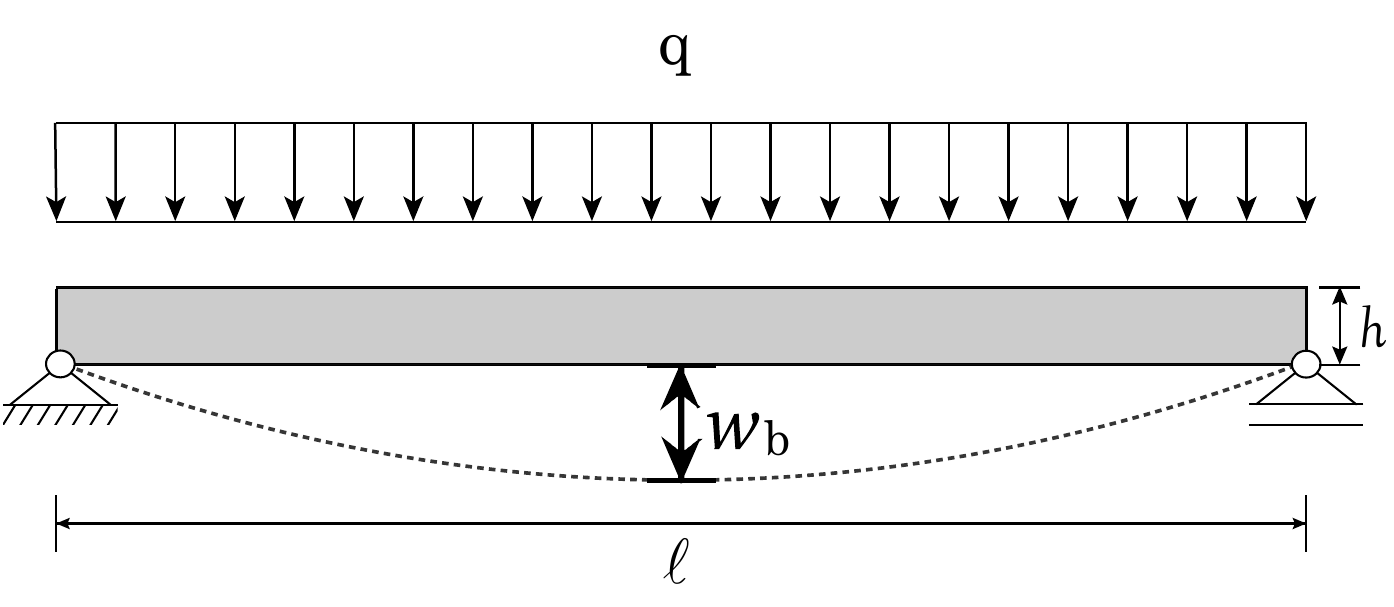}
         \label{fig:beam}
    }
    \caption{A truss structure with 23 bars and 6 upper cord nodes, used as the high fidelity (a), and its simply supported beam low-fidelity equivalent (b).}
\end{figure}

In our second application, we aim to investigate the scalability of our method when applied to higher-dimensional problems.
For this purpose, we consider a problem of engineering interest, and precisely, an ideal truss model with $23$ bars and $6$ upper cord nodes, as shown in \Cref{fig:truss} (see \citet{BlatmanCras2008}). 
This ten-dimensional model serves as the high fidelity.

The HF truss structure has height $H$ and length $L$, here considered constant, with $H = 2$m and $L = 24$m.  
The truss consists of two types of bars: horizontal bars with cross-sectional area $A_1$ and Young's modulus $E_1$, and oblique bars with cross-sectional area $A_2$ and Young's modulus $E_2$.
The truss is loaded with six vertical loads $P_i$ applied on the upper cord nodes. 
The quantity of interest is the mid-span displacement of the truss, denoted as $w_\text t$. 
Here, $w_\text t$ is calculated using an in-house finite element model programmed in Matlab.
The geometrical and material properties of the truss members, as well as the loads, are modeled as random variables, with the distributions provided in \Cref{tab:truss}.
\begin{table}[h]
    \caption{Truss model -- Input variables and their distributions}
    \centering
    \begin{tabular}{@{}lccc@{}}
        \toprule
        Variable & Distribution & Mean & Standard deviation \\ \midrule
        $E_1$, $E_2$ [Pa] & Lognormal & $2.1 \cdot 10^{11}$ & $2.1 \cdot 10^{10}$ \\
        $A_1$ [m$^2$] & Lognormal & $2 \cdot 10^{-3}$ & $2 \cdot 10^{-4}$ \\
        $A_2$ [m$^2$] & Lognormal & $1 \cdot 10^{-3}$ & $1 \cdot 10^{-4}$ \\
        $P_1$-$P_6$ [N] & Gumbel & $5 \cdot 10^{4}$ & $7.5 \cdot 10^{3}$ \\ \bottomrule
    \end{tabular}
    \label{tab:truss}
\end{table}

In real engineering applications, the truss displacement $w_\text t$ is commonly measured by laser measuring tools, and typically, such devices report a margin of error of $0.0015$~m ($1.5$~mm). 
Therefore, in this example, we add artificial noise $\varepsilon_{\text{H}} \sim \mathcal{N}(0,\, 0.0015)$ on the displacement $w_\text t$. 
To give a perspective on the level of the noise $\varepsilon_{\text{H}}$ with respect to $w_\text t$, the standard deviation of $w_\text t$, as computed from a reference PCE trained on $1,000$ data points (see, e.g., \citet{BlatmanJCP2011}), is $\hat \sigma_{w_\text t} \approx 0.0128$~m. This means that $ \sigma_{\varepsilon_\text H} \approx 0.12 \, \hat \sigma_{w_\text t}$.

We now consider as the LF counterpart of the described HF truss model a homogeneous simply supported beam with length $L$ and height $h$ subjected to a uniform loading $q$, as shown in \Cref{fig:beam}.
The mid-span deflection of the beam $w_\text{b}$ can be computed as the sum of deflections due to bending and shear. For slender beams with $h \ll L$, we can neglect the shear contribution and approximate $w_{b}$ as the deflection solely due to bending. Then,
\begin{equation}
    w_\text{b} = \frac{5\,q\,L^4}{384\,E\,I}.
\end{equation}
Here, we consider $q = \frac{\sum_{i=1}^6 P_i}{\ell}$. Moreover, the bending stiffness $EI$ of the beam is determined by the Young’s modulus $E$ (material property) and the moment of inertia $I$ (geometrical property). We consider $E = E_1$, and assuming that only the cords contribute to $I$, we can approximate it as $I = 2A_1 (\frac{H}{2})^2$.
Thus, the mid-span deflection of the LF beam model can be computed as:
\begin{equation}
    w_\text{b} = \frac{5\, L^3 \,  \sum_{i=1}^6 P_i}{384 \, E_1 \, 2A_1 \, \left(\frac{H}{2}\right)^2},
\end{equation}
where $H$ is the height of the corresponding HF truss.

For a data set of $1,000$ HF and LF samples, the Pearson correlation coefficient between the HF and LF data is 0.82, while the NRMSE is 1.40.

\subsubsection{MFSM performance and convergence}
Similarly to the previous analytical example, we now investigate the performance of our MFSM and its convergence with respect to the noise-free underlying HF truss model by computing the validation error $\epsilon_\text{val}$ for increasing HF experimental design size. 
More precisely, the HF ED size varies from 5 to 160 data points contaminated with the noise following the prescribed distribution, while the LF ED is fixed in all experiments to 300 data points. The larger amount of LF training data used compared to the previous application is due to the higher dimensionality and complexity of this application.
Again, for the computation of $\epsilon_\text{val}$, we use $N_{test} = 10^5$ noise-free HF data points generated using Latin Hypercube Sampling in the input space.
For the choice of basis in all PCEs involved in this example, we use degree adaptivity from degree 1 up to degree 8.

\begin{figure}[h]
    \centering
    \includegraphics[width=.6\textwidth, trim = 100 25 100 25]{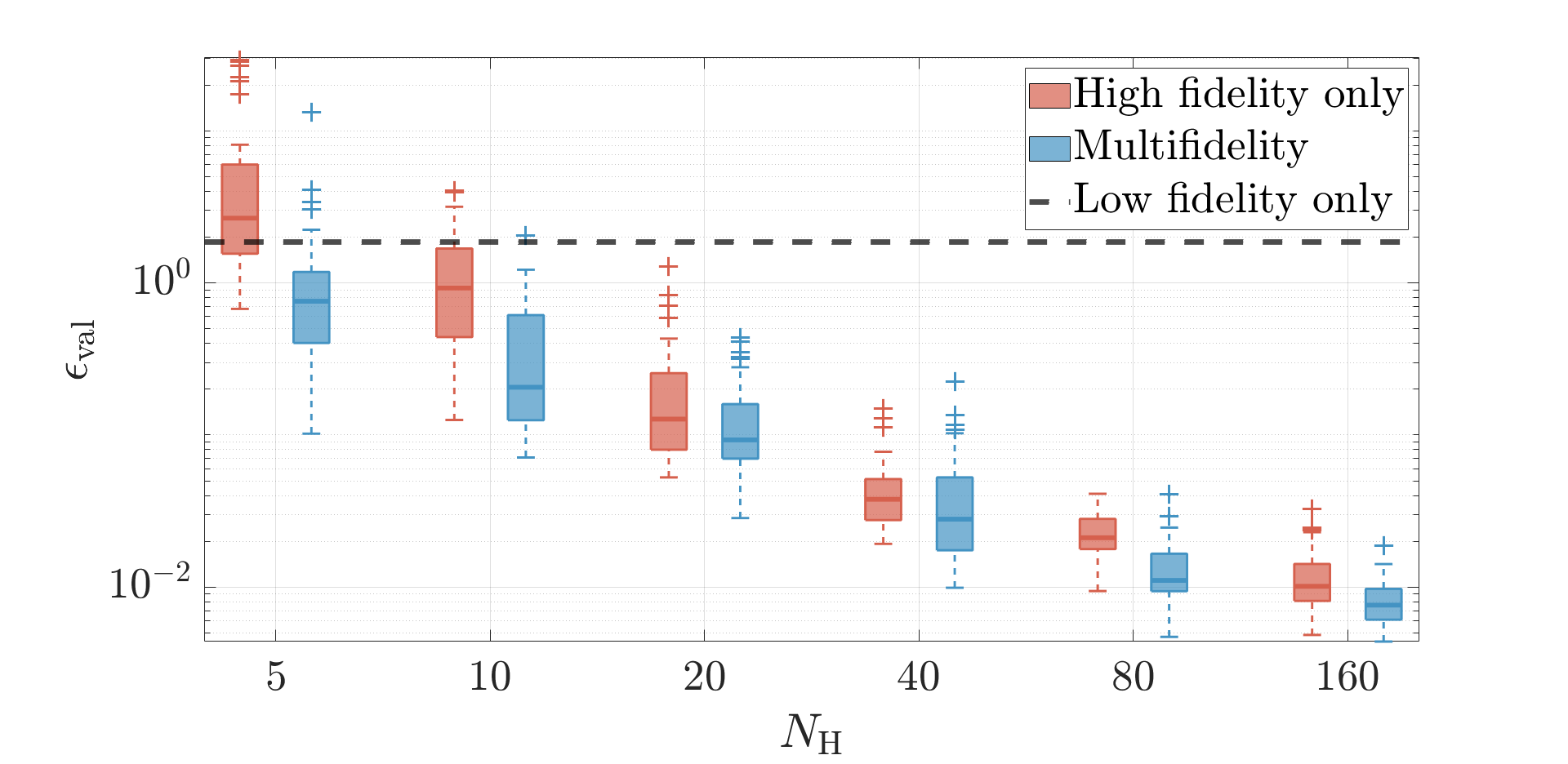}
    \caption{Truss model -- Convergence of the validation error $\epsilon_\text{val}$ for increasing number of HF training data. Comparison of our MFSM with a PCE model trained on HF data only. The dashed lines are the corresponding errors of a PCE model trained on LF data only.}
    \label{fig:truss_val_error}
\end{figure}

From \Cref{fig:truss_val_error}, we observe that our MFSM outperforms both the PCE model trained on HF data only, and the PCE model trained solely on LF data, across all the considered HF ED sizes. The performance difference between the MFSM and the HF PCE model is more pronounced when the available HF data comprises fewer than 20 data points.

\subsubsection{Confidence and prediction intervals}
We now proceed to the construction and evaluation of CIs and PIs for our MF truss model.
We set the HF ED size equal to 80 samples, as the previous study demonstrated satisfactory performance  at this sample size ($\epsilon_\text{val} \approx 1\%$), with only marginal improvement observed when doubling the sample amount. 
The HF data is contaminated with noise $\varepsilon_\text H \sim \cn(0, 0.0015)$. 
Furthermore, the LF ED consists of 300 samples.
Similarly to the previous application, $N_\text B = 1,000$ bootstrap replications are performed for the construction of the CIs and PIs.
Also, for the computation of the evaluation metrics reported in \Cref{tab:truss_CI_PI_eval}, $N_\text{rep} = 10$ replications are performed, and the test sets for the CI and PI evaluation consist of $N_\text t = 10,000$ data points each, obtained as described in the previous example.

From \Cref{tab:truss_CI_PI_eval}, we observe that for all the nominal coverage levels examined, our method provides reliable CIs and PIs. More precisely, the PI average coverage error ranges from less than $1\%$ to approximately $4\%$, while the corresponding error for the CIs varies from $2\%$ to $9\%$.
The observed error is always due to over-coverage, and though not ideal, is preferable to under-coverage and considered acceptable. 
Moreover, the PIs exhibit again closer coverage to the nominal levels compared to the CIs.

\begin{table}[h]
\centering
    \caption{Truss model -- Confidence and prediction intervals evaluation, where $\varepsilon_\text H \sim \cn(0, 0.0015)$, $N_\text H = 80, \, \text{and } N_\text L = 300$}
    \begin{tabular}{@{}lcccc@{}}
        \toprule
        $1-2\alpha\:$ & MCICP &  $\text{ACE}_\text{CI}$ & MPICP &  $\text{ACE}_\text{PI}$ \\ 
        \midrule
        $0.1$ & $0.124$ & $0.024$ & $0.109$ & $0.009$ \\
        $0.5$ & $0.593$ & $0.093$ & $0.542$ & $0.042$ \\
        $0.9$ & $ 0.959$ & $0.059$ & $0.935$ & $0.035$ \\
        $0.95$ & $0.986$ & $0.036$ & $0.973$ & $0.023$ \\
        \bottomrule
    \end{tabular}
    \label{tab:truss_CI_PI_eval}
\end{table}

\begin{figure}[h]
     \centering
     \subfigure[Selected dimension: Young's modulus $E_1$]{
         \includegraphics[width=.475\textwidth, trim = 25 0 10 0]{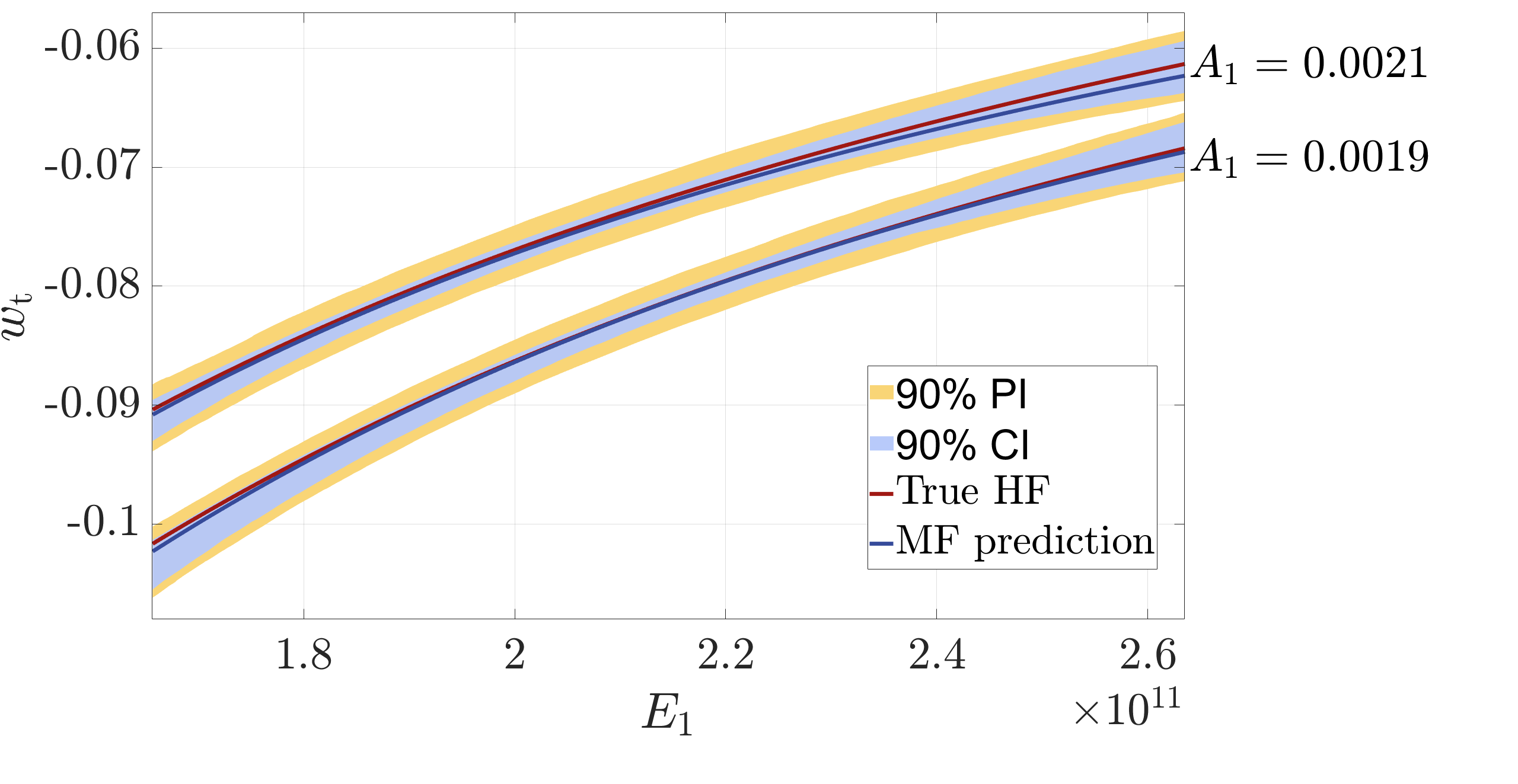}
         \label{fig:truss_E1vsA1}
     }%
    \hfill
    \subfigure[Selected dimension: cross-sectional area $A_1$]{
         \includegraphics[width=.475\textwidth, trim = 25 0 10 0]{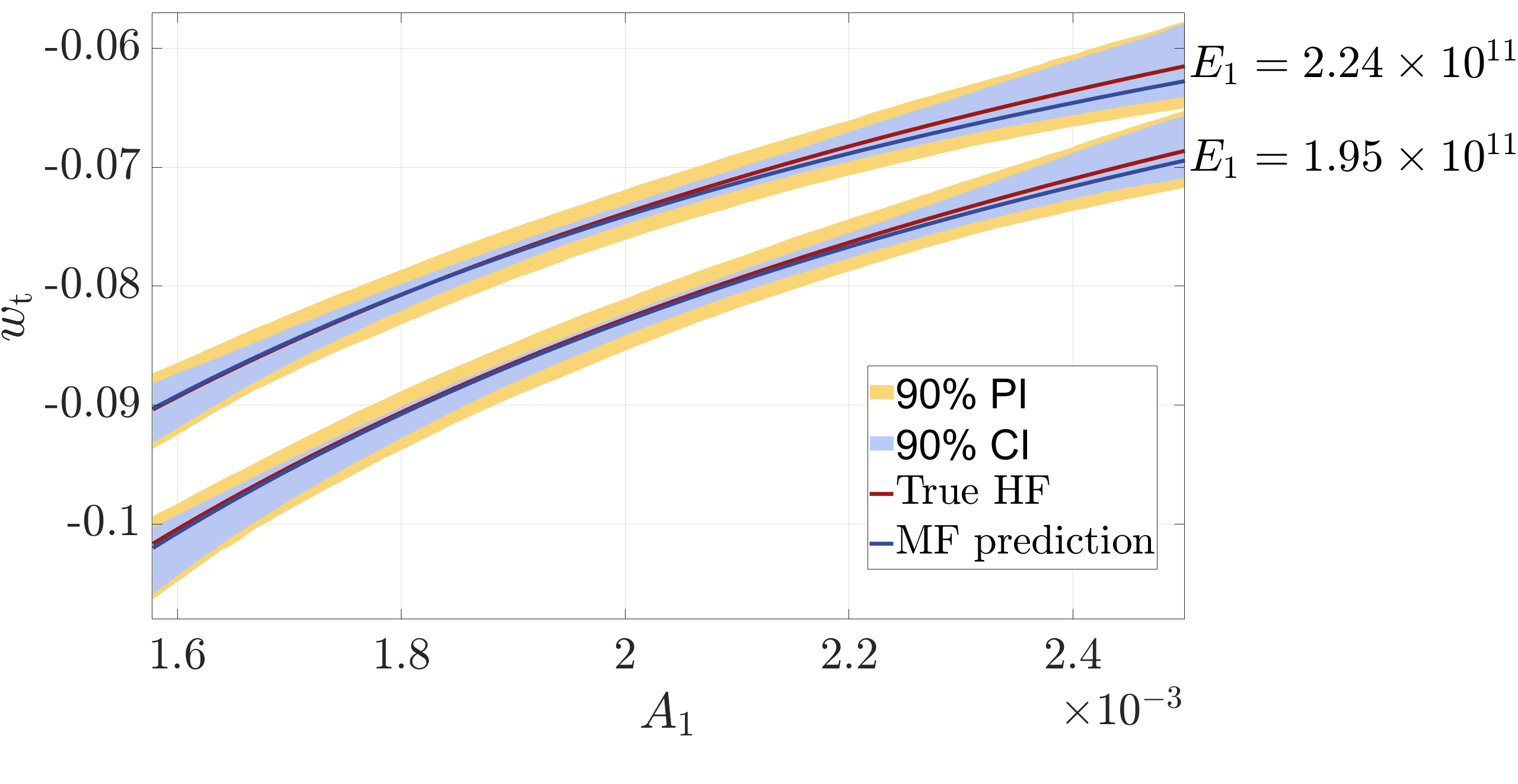}
         \label{fig:truss_A1vsE1}
    }

    \caption{Truss model -- $90\%$ confidence and prediction intervals along slices in the two selected dimensions for the MFSM trained on $80$ HF and $300$ LF data points. In each subplot, the blue area corresponds to the $90\%$ CI, the yellow area to the $90\%$ PI, while the red and blue lines depict the true noise-free HF response and the MFSM prediction respectively.}
    \label{fig:truss_CIs_PIs}
\end{figure}

To illustrate the constructed CIs and PIs in this application, we select the random variables $E_1$ and $A_1$ as the most important ones, according to a sensitivity analysis on the HF truss model performed by \citet{BlatmanJCP2011}. 
\Cref{fig:truss_CIs_PIs} illustrates the $90\%$ CIs (blue area) and PIs (yellow area) along slices in the two selected dimensions (with all the other parameters kept at their mean values), as well as the true HF model response and our MFSM response (red and blue line respectively) for the selected HF and LF experimental designs. 
\Cref{fig:truss_E1vsA1} shows these quantities as a function of $E_1$ for two different values of $A_1$ that correspond to its $0.25$- and $0.75$-quantiles, while the rest input random variables are fixed at their mean. 
Similarly, \Cref{fig:truss_A1vsE1} shows the same quantities as a function of $A_1$ for two different values of $E_1$.

\subsection{Real-world application: aero-servo-elastic simulation of a wind turbine}
\label{sec:appl_wind_turb}
In our last application, we aim to explore the applicability and performance of our framework in a real-world application involving real wind turbine simulations, performed by \citet{AbdallahPEM2019}. For this study, an onshore wind turbine standing on a $90$~m tower is considered, with a rotor diameter of $110$~m and a rated power of $2$~MW. 
The goal is to investigate the impact of the wind speed, turbulence intensity, and shear profile on the variation of the extreme loads, and specifically the maximum flapwise bending moment at the wind turbine blade root. The wind speed, turbulence intensity, and shear exponent are modeled as random variables, whose distributions are described in \Cref{tab:wind_turb_distr}.

\begin{table}[h]
\centering
    \caption{Wind turbine simulations -- Input variables and their distributions}
    \begin{tabular}{@{}lcc@{}}
        \toprule
        Variable & Distribution & Parameters \\ \midrule
        Wind speed ($U$) [m/s] & Uniform & $[4, 25]$\\
        Turbulence intensity ($\sigma_\text U$) [m/s] & Uniform & $[0.1, 6]$ \\
        Wind shear exponent ($\alpha$) [-] & Uniform & $[-1, 1.5]$ \\ \bottomrule
    \end{tabular}
    \label{tab:wind_turb_distr}
\end{table}

\citet{AbdallahPEM2019} used two different numerical aeroservo-elastic simulators, namely Bladed and FAST. Simulation data from Bladed are considered as the high-fidelity data, while the FAST simulation data are considered to be the low-fidelity data. Details on the technical characteristics of the Bladed and FAST simulators can be found in \citet{AbdallahPEM2019,bossanyi2003gh,Jonkman2005}. 
A 10-minute time series simulation in FAST takes about 5 minutes to run in real time, whereas the same simulation in Bladed takes approximately 30 minutes. Consequently, there are fewer Bladed simulations compared to FAST simulations. 
The experimental designs for the Bladed and FAST simulations can be found in \Cref{tab:simulations_ED}.

\begin{table}[h]
    \caption{Experimental design for Bladed and FAST simulations}
    \centering
    \begin{tabular}{@{}lccc@{}}
        \toprule
        Simulator & Wind speed ($U$) & Turbulence intensity ($\sigma_\text U$) & Wind shear exponent ($\alpha$) \\ 
        \midrule
        Bladed & $4, 8, 10, 12, 15, 20, 25$ & $0.1, 1, 2, 3, 4, 5, 6$ & $\pm 1, \pm 0.6, \pm 0.2, \pm 0.1, 0, 1.5$ \\
        FAST & $4,\, 5,\, 6,\, ...,\, 25$ & $0.1, 1, 2, 3, 4, 5, 6$ & $\pm 1, \pm 0.6, \pm 0.2, \pm 0.1, 0, 1.5$ \\ \bottomrule
    \end{tabular}
    \label{tab:simulations_ED}
\end{table}

Each combination of wind speed, turbulence intensity, and shear exponent was used to generate wind fields. A wind field was used as input to the wind turbine simulators to produce a time series realization of the structural response.  Due to the stochasticity in the wind field generation, $12$ different stochastic seeds were used for Bladed and $24$ different seeds for FAST. Excluding certain combinations of input parameters that are not realistic resulted in $4,344$ and $33,480$ simulations for Bladed and FAST respectively. 
Our HF data set is the mean of the output maximum flapwise bending moment at the wind turbine blade root over the $12$ time series from Bladed, thus $362$ data points.
Moreover, our LF data set is the mean system response over the $24$ time series from FAST, resulting in a total of $1,395$ data points.
There are 261 common input samples between the HF and LF data sets. For these samples, the Pearson correlation coefficient between the HF and LF data is 0.95, with a NRMSE of 0.38.

Both FAST and Bladed are deterministic simulators, meaning that repeated runs with the same initial conditions and same input wind field produce the same output time series. However, since the wind fields are stochastic with respect to the three input random variables, both the HF and the LF data coming from the wind turbine simulators contain stochasticity. We treat this stochasticity as additive homoscedastic noise, and our MF framework remains applicable as is.

\subsubsection{MFSM performance and convergence}
In this application, we explore the performance of our MFSM for increasing HF experimental design size, equal to $10\%, 20\%,..., 70\%$ of the total HF data available, while keeping the LF ED fixed to all the available LF data. 
We compute the validation error $\epsilon_\text{val}$ of the MFSM on a test set consisting of the $30\%$ of the HF data which was not used for training any of the MFSMs at each given replication:  $N_\text{test} = 0.3\times 362 = 109$ data points.
For the choice of basis in all PCEs involved in this application, we use degree adaptivity from degree 1 up to degree 10.

\begin{figure}[h]
    \centering
    \includegraphics[width=.6\textwidth, trim = 100 25 100 25]{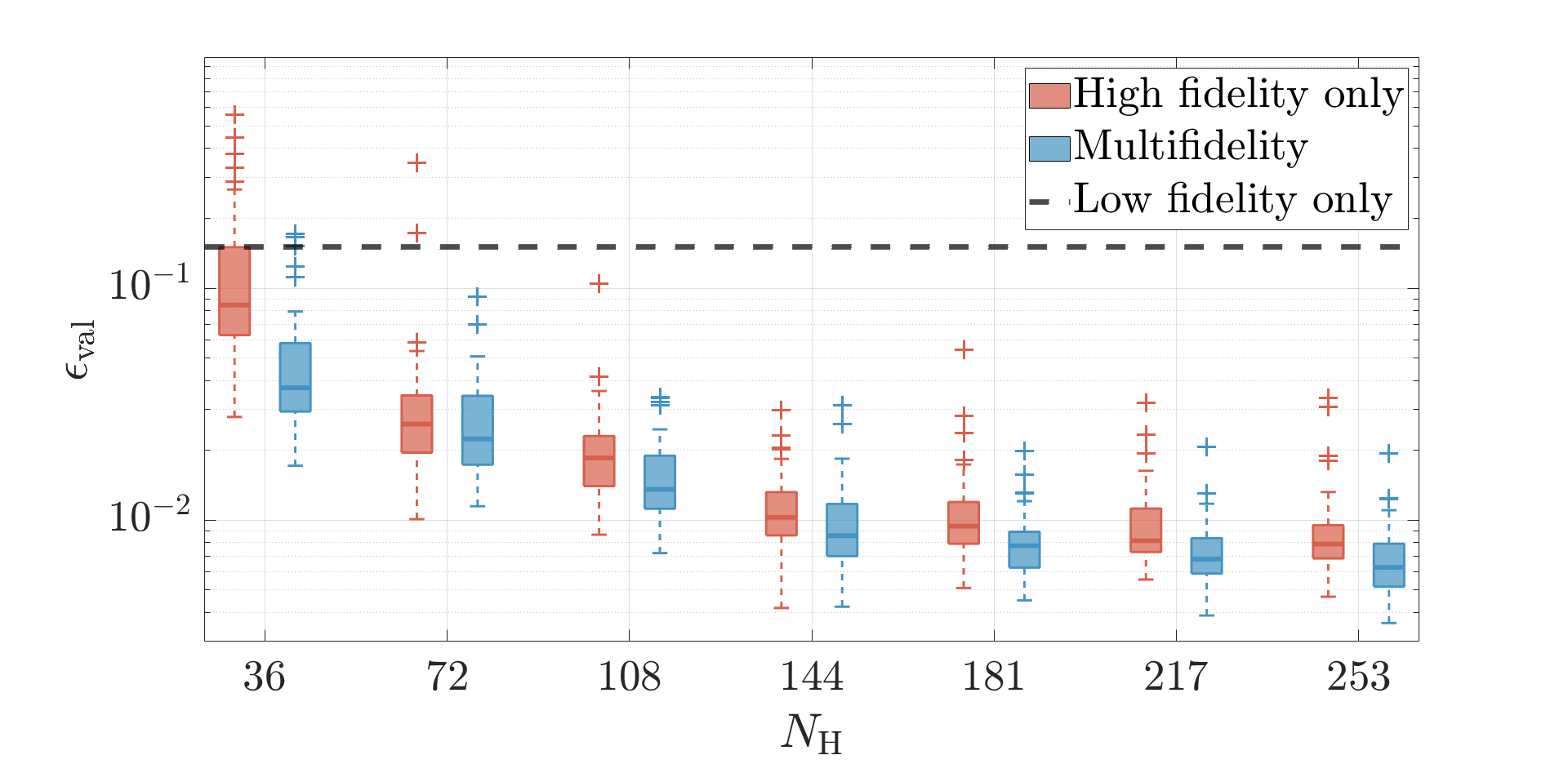}
    \caption{Wind turbine application -- Convergence of the validation error $\epsilon_\text{val}$ for increasing number of HF training data. Comparison of our MFSM with a PCE model trained on HF data only. The dashed lines are the corresponding errors of a PCE model trained on LF data only.}
    \label{fig:wind_turb_val_error}
\end{figure}

As shown in \Cref{fig:wind_turb_val_error}, our MFSM outperforms the PCE model trained on HF data only, the difference being more evident especially for small HF EDs. 
We can notice that we achieve satisfactory performance ($\epsilon_\text{val} < 1\%$) already for $N_\text H = 144$, which corresponds to $40\%$ of the available HF data. 

Let us note that in this application, $\epsilon_\text{val}$ is not expected to approach zero as the HF ED size increases, because $\epsilon_\text{val}$ is now computed with respect to noisy HF data. Instead,  $\epsilon_\text{val}$ is expected to converge to a value representative of the noise present in the HF data.

\subsubsection{Confidence and prediction intervals}
In this section, we provide the CIs and PIs for our MFSM prediction. 
In this real-world application, the true noise-free HF function is unknown, limiting our ability to assess the reliability of the constructed CIs. 
Thus, we are able to only appraise the constructed PIs on a test set from the HF data. 
We use $70\%$ of the available HF data for training and the rest $30\%$ for the PI evaluation, and we perform $N_\text{rep} = 10$ replications using different seeds to account for the statistical uncertainty in the HF random design and the bootstrap sampling.

From \Cref{tab:wind_turb_PI_eval}, we observe that the coverage of our PIs is close to the nominal, albeit generally overestimated.
Once again in this application, the errors are due to over-coverage, which is more evident for nominal coverage $50\%$ and $90\%$.

\begin{table}[h]
\centering
    \caption{Wind turbine application -- Prediction intervals evaluation}
    \begin{tabular}{@{}lcc@{}}
        \toprule
        $1-2\alpha\:$ & MPICP &  $\text{ACE}_\text{PI}$ \\ 
        \midrule
        $0.1$ & $0.119$ & $0.019$ \\
        $0.5$ & $0.583$ & $0.083$ \\
        $0.9$ & $0.950$ & $0.050$ \\
        $0.95$ & $0.981$ & $0.031$ \\
        \bottomrule
    \end{tabular}
    \label{tab:wind_turb_PI_eval}
\end{table}

\begin{figure}[h]
     \centering
     \subfigure[Selected dimension: wind speed $U$]{
         \includegraphics[width=.475\textwidth, trim = 25 0 150 0]{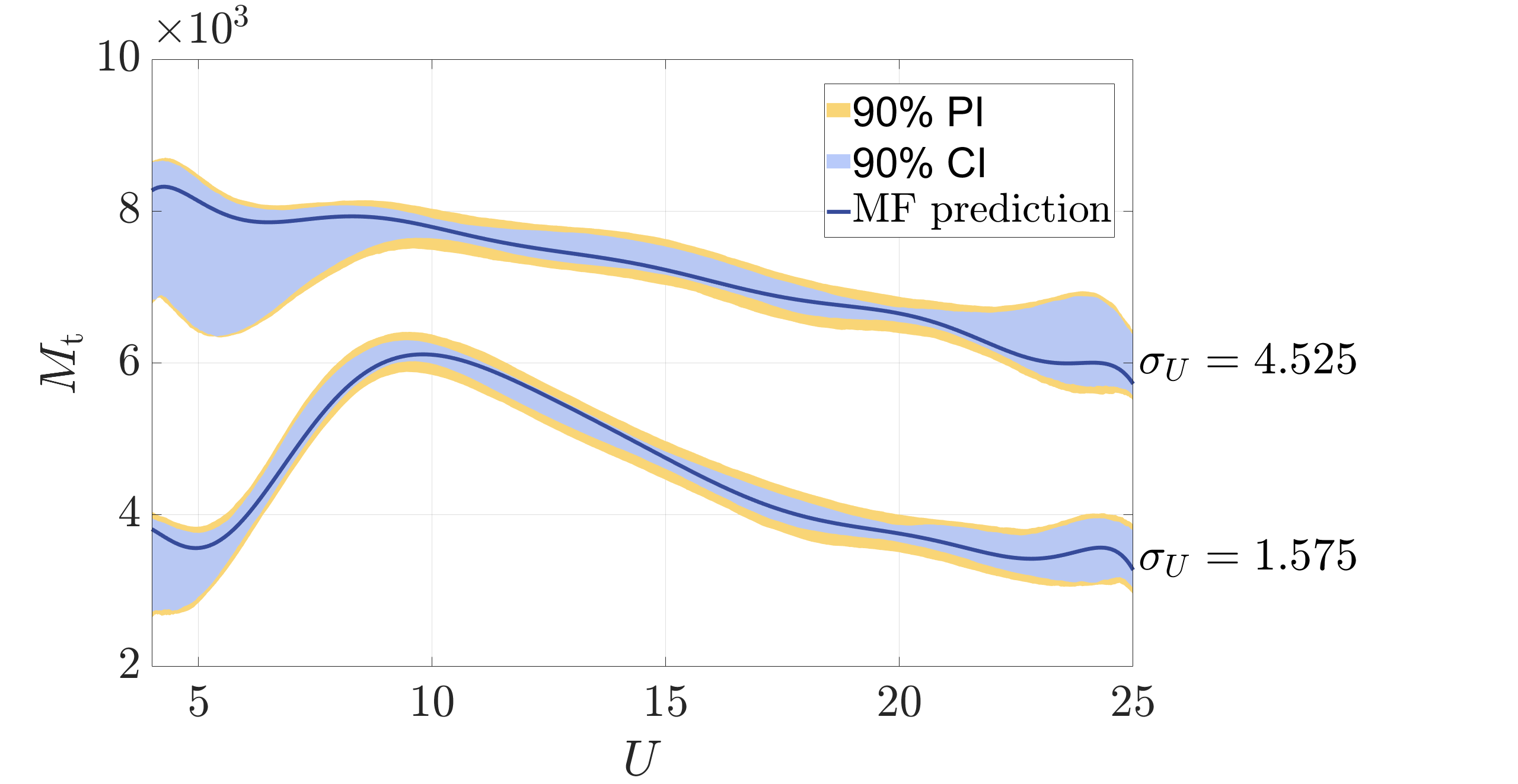}
         \label{fig:wind_turb_Uvssigma}
     }%
    \hfill
    \subfigure[Selected dimension: turbulence intensity $\sigma_\text U$]{
         \includegraphics[width=.475\textwidth, trim = 25 0 150 0]{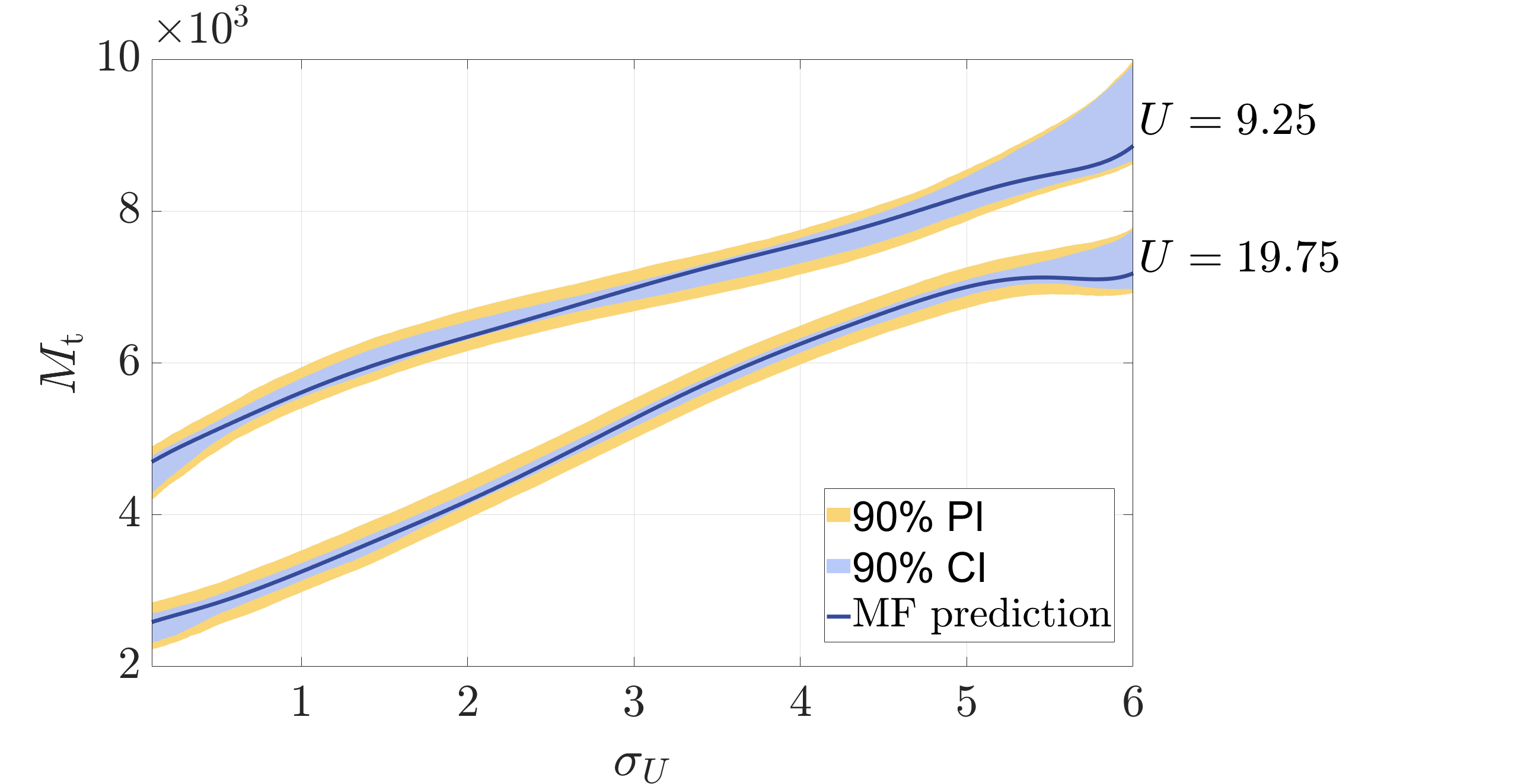}
         \label{fig:wind_turb_sigmaVsU}
    }

    \caption{Wind turbine application -- $90\%$ confidence and prediction intervals along slices in the two selected dimensions for the MFSM trained on $253$ HF and $1,395$ LF data points. In each subplot, the blue area corresponds to the $90\%$ CI, the yellow area to the $90\%$ PI, while the blue line depicts the MFSM prediction respectively.}
    \label{fig:wind_turb_CIs_PIs}
\end{figure}

The predicted extreme flapwise bending moment as well as the $90\%$ CIs and PIs in the wind speed and turbulence intensity dimensions are illustrated in \Cref{fig:wind_turb_CIs_PIs}.
More precisely, \Cref{fig:wind_turb_Uvssigma} shows the MFSM prediction and the corresponding intervals as a function of the wind speed $U$ for two different values of the turbulence intensity $\sigma_\text U$ that correspond to its $0.25$- and $0.75$-quantiles, while the wind shear exponent $\alpha$ is fixed at its mean. 
Similarly, \Cref{fig:wind_turb_sigmaVsU} depicts the MFSM prediction and the corresponding intervals as a function of $\sigma_\text U$ for two different values of $U$.

\section{Conclusions}
\label{sec:conclusions}
In this paper, we presented a novel and comprehensive framework for multi-fidelity surrogate modeling that effectively handles noisy data and incorporates epistemic uncertainty arising from limited training information.
Our regression-based approach aims to emulate the assumed underlying noise-free HF model, and provides accurate predictions and denoising capabilities. It also offers uncertainty estimations with respect to not only the underlying HF model, but also to unseen noise-contaminated HF observations in the form of confidence and prediction intervals respectively, constructed using the bootstrap methodology. 
The proposed framework is applied to combine experimental data and computer simulations, where noisy measurements, considered as the high fidelity, are combined with white-box computer simulations, treated as the low fidelity.
However, our framework is not limited solely to this particular scenario. Its versatility extends to situations where both the HF and the LF components are experiments or simulations.

Our framework proves its efficacy in various scenarios, including a one-dimensional analytical example and a ten-dimensional application that incorporates a high-fidelity finite element model alongside a low-fidelity analytical approximation.
In both scenarios, noise was artificially added to the HF data. 
In these synthetic examples, our multi-fidelity surrogate modeling method clearly outperforms both surrogate models trained on the available high- and low-fidelity data separately, and it shows convergence to the noise-free HF model with increasing number of HF training data. 
Moreover, the constructed confidence and prediction intervals exhibit remarkably high reliability, achieving coverage close to the nominal levels.
Finally, our framework demonstrates its versatility and potential by being applied on a real-world example involving wind turbine simulations of different fidelity levels. In this application, our method provides again accurate predictions and reliable prediction intervals.

It should be noted that the reliability of the confidence and prediction intervals comes at the cost of computational time for their construction. 
For low-dimensional problems, this time can be considered negligible, but in higher dimensions ($\ge$~10) this is not the case anymore. 
Nonetheless, this time overhead is incurred only during the training phase and can subsequently be mitigated as the bootstrap results can be stored. Thus, any subsequent inference including predictions at unobserved points along with their associated confidence and prediction intervals, can be instantaneously accessed.

In our future work, we plan to extend our methodology in various directions.
Firstly, different types of surrogate models and data fusing methodologies can be explored for our multi-fidelity surrogate model construction.
Additionally, to improve the performance of confidence and prediction intervals, different techniques can be investigated, such as more advanced bootstrap methods \citep{efron1994introduction}.
Finally, the provided uncertainty estimations for the multi-fidelity model predictions can be employed for different purposes, one being the adaptive design of sampling strategies to obtain new samples from the high- and the low-fidelity models.

\section*{Declaration of Competing Interest}
The authors declare that they have no known competing financial interests or personal relationships that could
have appeared to influence the work reported in this paper.

\section*{Acknowledgments}
The authors express their sincere appreciation to Styfen Sch{\"a}r for his valuable contribution to the second of the three applications presented in this paper, particularly for formulating the low-fidelity simply supported beam approximation of the high-fidelity truss structure.

The research presented in this paper is part of the GREYDIENT project, funded by the European Union’s  Horizon 2020 research and innovation
program under the Marie Sk\l{}odowska-Curie 
grant agreement No 955393. GREYDIENT's support is gratefully acknowledged.

\bibliography{bibliography.bib}

\end{document}